\documentclass[12pt]{article}
\usepackage{graphicx}
\usepackage{latexsym}
\title{Time and probability: From classical mechanics to relativistic Bohmian mechanics}
\author{Hrvoje Nikoli\'c \\
Theoretical Physics Division, Rudjer Bo\v{s}kovi\'{c} Institute, \\
P.O.B. 180, HR-10002 Zagreb, Croatia \\
{\normalsize e-mail: hnikolic@irb.hr} \\
\makebox[1in]{} \\
}
\date{\today}
\begin{document}
\maketitle
\begin{abstract}
Bohmian mechanics can be generalized to a relativistic
theory without preferred foliation, with a price of 
introducing a puzzling concept of spacetime probability
conserved in a scalar time. We explain how analogous concept 
appears naturally in classical statistical mechanics 
of relativistic particles, with scalar time being identified
with the proper time along particle trajectories.
The conceptual understanding of relativistic Bohmian mechanics 
is significantly enriched by this classical insight. 
In particular, the analogy between classical and Bohmian mechanics
suggests the interpretation of Bohmian scalar time as 
a quantum proper time different from the classical one, 
the two being related by a nonlocal scale factor calculated
from the wave function. 
In many cases of practical interest, including the macroscopic measuring apparatus,
the fundamental spacetime probability explains 
the more familiar space probability as an emergent approximate description.
Requiring that the quantum proper time
in the classical limit should reduce to the classical proper time,
we propose that only massive particles have Bohmian trajectories.
An analysis of the macroscopic measuring apparatus made up of 
massive particles restores agreement with 
the predictions of standard quantum theory.
\end{abstract}

\noindent
PACS: 03.65.Ta, 05.20.-y

\newpage

\tableofcontents

\newpage

\section{Introduction}
\label{SEC1} 

\subsection{Motivation}

The Bohmian formulation of non-relativistic quantum mechanics (QM) \cite{bohm}
is the best known and most successful hidden-variable formulation of quantum theory 
\cite{book-bohm,book-hol,book-durr,apl-book}.
However, the explicit nonlocality of the theory causes difficulties in attempts
to reconcile it with the theory of relativity.
In fact, contrary to a wide belief, it is not so difficult to write down nonlocal 
relativistic-covariant equations for particle trajectories,
without introducing any preferred foliation of spacetime 
\cite{berndl96,nikFPL05,nikAIP06}. The problems are (i) to make unambiguous probabilistic 
predictions resulting from such a theory, and (ii) to show that they are compatible 
with the probabilistic predictions of standard QM. 
In the simplest case of a single relativistic particle without spin, 
the problem can be reduced to the fact that the Bohmian 3-velocity $v^i$ 
does {\em not} obey a space-equivariance equation of the form
\begin{equation}\label{e1}
 \frac{\partial|\psi|^2}{\partial t} + \partial_i(|\psi|^2 v^i) =0 ,
\end{equation}
where $\psi({\bf x},t)$ is the wave function, ${\bf x}=(x^1,x^2,x^3)$,
$\partial_i=\partial/\partial x^i$, $i=1,2,3$, and the summation over 
repeated index $i$ is understood.
The violation of (\ref{e1}) implies that $|\psi({\bf x},t)|^2$ cannot 
be the probability density $\rho({\bf x};t)$ that the particle
will have the position ${\bf x}$ at time $t$.

As a reaction to this problem, it has been suggested several times 
\cite{book-bohm,fanchi,nikFP09,nikIJQI09,nikIJMPA10,hernandez,nikIJQI11,nikFP12,nikBOOK12}
that (\ref{e1}) should be replaced by a manifestly covariant spacetime-equivariance equation
\begin{equation}\label{e2}
 \frac{\partial|\psi|^2}{\partial s} + \partial_{\mu}(|\psi|^2 V^{\mu}) =0 ,
\end{equation} 
where $\mu=0,1,2,3$ is the spacetime vector index and $s$ is a scalar parameter
used to parameterize particle trajectories $X^{\mu}(s)$ with velocities
\begin{equation}\label{e3}
 \frac{dX^{\mu}}{ds}=V^{\mu} . 
\end{equation}
The spacetime-equivariance equation provides the consistency of the spacetime 
probability density $\rho=|\psi|^2$, where the sum of all probabilities is
\begin{equation}\label{e4}
 \int d^4x \, |\psi|^2 =1 .
\end{equation}

However, the covariant approach based on (\ref{e2})-(\ref{e4}) leads to several
conceptual difficulties. Standard QM is interpreted in terms of space probability density,
so how can it be compatible with the spacetime probability density?
Isn't the left-hand side of (\ref{e4}) actually infinite?
If, as the concept of spacetime probability suggests, $x^0$ may be uncertain,
then why in physics time $t$ can usually be viewed as certain?
What is the physical meaning of the parameter $s$?
How the measurable probabilistic predictions of standard QM can be reproduced
from the spacetime probability density?
The results in the literature provide only partial answers to these questions
(see e.g. \cite{nikBOOK12} for a review). 
The main goal of the present paper is to give a more thorough 
answer to these and other related questions.

\subsection{Main ideas}

To understand the main idea, let us first reformulate the problem as follows.
Suppose that one wants to interpret $|\psi|^2$ as probability density conserved in 
time. But the theory of relativity asserts that the time coordinate is not unique.
So the question is: {\em which} time?

In the standard Copenhagen interpretation of QM, one is inclined to say that time is simply 
the time of the observer who attributes the probabilistic interpretation to $|\psi|^2$.
But the observer in Copenhagen interpretation is an {\em external} observer,
not himself described by QM. On the other hand, one of the goals of Bohmian mechanics
is precisely to remove the external observer from the theory. So how to specify time 
without referring to an external observer?

One possibility is to introduce a preferred foliation of spacetime,
which defines time in terms of spacelike hypersurfaces of constant time.
To make it relativistic in spirit, the foliation may not be given {\it a priori},
but somehow determined dynamically from the wave function 
\cite{durr99,goldstein-zanghi,nik_fol,durr13}. Unfortunately, the existing
concrete proposals of that kind do not seem very satisfying. 
So, can time be specified without preferred foliation and
without an external observer?

Now our central conceptual idea is the following. 
Bohmian mechanics is a fundamentally deterministic theory, so any
probabilistic aspect of it must necessarily be associated with a lack of knowledge.
Therefore, there must be a subject which lacks this knowledge.
But there should not be any external subject, so the only remaining possibility
is an {\em internal} subject. In other words, the probability 
must be associated with knowledge available (or unavailable)  
{\em to the particles themselves}. (A human observer is just a special case,
as humans, after all, are also made of particles.) 
Therefore, {\em the physical time with respect to which the probability is conserved
must be the time experienced by the particles}. 

This simple but crucial idea represents the core idea on which this paper is based.
In particular, in classical relativistic mechanics it leads immediately
to the result that probability must be conserved with respect to the
{\em proper time} associated with particle trajectories. We shall see that 
this also explains why $x^0$ may not be certain in the relativistic case, while
it must be certain in the non-relativistic limit.

However, it will turn out that the identification of $s$ with proper time 
along particle trajectories is more subtle than it may seem at the first sight.
The classical proper time $\tau$ along a particle trajectory is given by 
\begin{equation}\label{e5}
 d\tau^2=\eta_{\mu\nu}dX^{\mu}dX^{\nu} ,
\end{equation}
where $\eta_{\mu\nu}={\rm diag}(1,-1,-1,-1)$ is the Minkowski metric, and we work in units
in which the velocity of light is $c=1$. This relation is completely local, in the sense 
that proper time experienced by some particle does not depend on the motion of other
particles. In the quantum case it will turn out that (\ref{e5}) must be modified
with a nonlocal expression.
If we have $n$ particles with trajectories $X_a^{\mu}(s)$, $a=1,\ldots,n$, 
then instead of (\ref{e5}) we will have
\begin{equation}\label{e6}
 ds^2=f_a(X_1,\ldots,X_n) \, \eta_{\mu\nu}dX_a^{\mu}dX_a^{\nu} ,
\end{equation}
where $d\tau_a^2=\eta_{\mu\nu}dX_a^{\mu}dX_a^{\nu}$ is the classical local proper time,
and $f_a(X_1,\ldots,X_n)$ is a nonlocal function determined by the 
nonlocal many-particle wave function. 

At first sight, the appearance of a nonlocal scale factor $f_a$ in (\ref{e6}) may seem
rather puzzling, but in fact it is not completely new. In \cite{nikFP12} $f_a$ has been proposed
to be proportional to $|\psi|^4$. In this paper we shall use the analogy with
classical relativistic mechanics, non-relativistic Bohmian mechanics, and even
non-relativistic classical mechanics, to motivate a much more natural choice
of $f_a$. In particular, we shall see that an analogue of (\ref{e6}) 
with a nonlocal function $f_a$ appears even in non-relativistic Bohmian mechanics.    

\subsection{Organization of the paper}

This paper is rather long. The main new results could certainly be exposed
in a shorter one. But the ambition of the present paper is more than that.
We want this paper to be a fully comprehensive pedagogic self-contained
exposition of relativistic Bohmian mechanics, containing not only the
results which are truly new, but also some older, 
and yet not-widely-known results 
which, in our opinion, are necessary for thorough understanding.
In particular, since the many-time formalism is essential for the
relativistic covariant formulation of Bohmian mechanics, we want to be sure
that readers fully grasp it. For that purpose, we spend a lot of space to explain
in fair detail how the many-time formalism appears in more familiar contexts
such as relativistic classical mechanics, non-relativistic quantum 
mechanics, and even non-relativistic classical mechanics.

For impatient readers who are already familiar with the many-time formalism
and with some of the results in 
\cite{nikFP09,nikIJQI09,nikIJMPA10,hernandez,nikIJQI11,nikFP12,nikBOOK12} 
(but feel that something is still missing there), the quickest route to grasp the main new
results of this paper is as follows. First, to understand why in relativity
one needs to introduce spacetime probability conserved in proper time,
one should read Secs.~\ref{SEC2.1} and \ref{SEC2.1'}. Second, 
for our novel perspective on the the concept of proper time, one should see
Secs.~\ref{SEC3.4} and \ref{SEC4.2}. After that, one is ready to grasp the main 
general principles of relativistic Bohmian mechanics through 
Secs.~\ref{SEC5.2}, \ref{SEC5.3} and \ref{SEC5.4}.
Finally, to see how relativistic Bohmian mechanics reproduces 
the measurable predictions of standard QM, one should not miss Sec.~\ref{SEC7.2}.

For all others, who want to get slowly and don't want to miss any details,
the paper as a whole is organized as follows.

We start with some conceptual preliminaries in Sec.~\ref{SEC2}, 
explained very simply from the technical point of view.
Yet, from the conceptual point of view, this section is essential
for proper understanding of the rest of paper, so its relevance
should not be underestimated. In particular, it explains why probability is conserved 
in time, why in the relativistic case this should be identified with the 
{\em proper time}, and how that leads to spacetime probability in the 
relativistic case.

Sec.~\ref{SEC3} is devoted to non-relativistic mechanics (both classical and
quantum), including a probabilistic approach based on space equivariance,
with an emphasis on the relation between the many-time formulation on one side, 
and the standard single-time formulation on the other.
In addition, we outline how 
Newton time can be viewed in a novel way as a non-relativistic
proper time associated with the particle trajectory.
(The many-time formulation
of non-relativistic QM can also be found elsewhere. See e.g. 
\cite{dirac,rosenfeld,dfp,bloch,tomon}
for early papers, or \cite{tumulka} and references therein for more recent work.
Some aspects of the many-time formulation
of classical mechanics are discussed in \cite{longhi,tumulka2}.)

Sec.~\ref{SEC4} is devoted to classical relativistic mechanics, generalizing the
corresponding non-relativistic results from Sec.~\ref{SEC3} in both 
the many-time and single-time forms. In particular, the non-relativistic space equivariance
generalizes to relativistic spacetime equivariance. We pay particular attention to explain in detail how, 
under certain circumstances, a space-equivariance equation can be derived from the more fundamental 
spacetime equivariance.

In Sec.~\ref{SEC5} we finally formulate the main principles of relativistic 
Bohmian mechanics, emphasizing the two fundamental postulates from which everything else
can be derived. In particular, we derive the spacetime-equivariance equation 
and show that, under a widely valid assumption, it can be approximated by a space-equivariance equation. 

In Sec.~\ref{SEC6} we briefly outline some generalizations of the theory, including particles with spin,
influence of classical background fields, and quantum field theory. We also discuss the case of
massless particles and argue that they do not have Bohmian trajectories. 

Sec.~\ref{SEC7} is devoted to the theory of quantum measurements, which is
essential to understand how relativistic Bohmian mechanics reproduces 
the measurable predictions of standard QM.

The conclusions are drawn in Sec.~\ref{SEC8}.

In Appendix \ref{SECA} we discuss how some arguments in the literature 
against existence of relativistic-covariant versions of Bohmian mechanics are circumvented 
by our theory, and argue that nonlocality of the quantum proper time 
cannot be avoided in a wide class of theories obeying
a spacetime-equivariance equation.  
In Appendix \ref{SECB} we derive some technical results concerning properties of positive-energy
wave functions.  
In Appendix \ref{SECC} we discuss the difference between macroscopic and microscopic proper time
in classical mechanics.

\section{Conceptual preliminaries}
\label{SEC2}

 \subsection{Probability and knowledge}
\label{SEC2.1}

In general, a probabilistic description is used when the values of some variables
are not known with certainty. Let there be $N$ such unknown variables
$u_1,\ldots,u_N$. The probability may also depend on some known variables
$k_1,\ldots,k_M$. For example, in QM the known variable is usually
time, while in statistical physics in thermodynamic equilibrium the known variables
include temperature and chemical potential. But whatever the known and unknown variables are,
the probabilistic description always involves the probability density
$\rho(u_1,\ldots,u_N;k_1,\ldots,k_M)$ satisfying
\begin{equation}\label{e2.1}
\int d^Nu \, \rho(u_1,\ldots,u_N;k_1,\ldots,k_M) =1 .  
\end{equation}
The known variables $k_1,\ldots,k_M$ are not described probabilistically, which is
why (\ref{e2.1}) does not involve an integration over $d^Mk$.
When the only known variable is time, one interprets (\ref{e2.1}) by saying that
probability is conserved in time. In particular, in standard QM the probability is conserved in
time due to unitarity. In general, we shall say that (\ref{e2.1}) means that
the probability is conserved in the variables $k_1,\ldots,k_M$.

Even though $k_1,\ldots,k_M$ are known, formally they can also be treated 
probabilistically, by using $\delta$-functions. 
If the variables $k_a$ are known to have some definite values 
$K_a$, then we may introduce the probability density
\begin{eqnarray}\label{e2.2}
 \rho(u_1,\ldots,u_N,k_1,\ldots,k_M) &=& \rho(u_1,\ldots,u_N;k_1,\ldots,k_M) \times
\nonumber \\ 
& &
\delta(k_1-K_1)\cdots \delta(k_M-K_M) .
\end{eqnarray}
It satisfies
\begin{equation}\label{e2.3}
\int\int d^Nu \, d^Mk\, \rho(u_1,\ldots,u_N,k_1,\ldots,k_M) =1 .  
\end{equation}

The probability depends on available knowledge. However, different levels of knowledge exist in physics, 
which means that one can attribute different probabilities
to the same variable in the same physical conditions, depending on which level of knowledge
one is having in mind. Roughly, the main levels of knowledge in physics can be classified as follows:
\begin{itemize}

\item Level I -- {\it Empirical knowledge}:
The knowledge obtained by measurement.

\item Level II -- {\it Theoretical knowledge}:
The knowledge obtained by theory. It can be further split into two sub-levels:

\begin{itemize}

\item Level II.A -- {\it Theoretical knowledge in practice}:
The knowledge determined by equations which are sufficiently simple
to solve in practice. For example, it may be a two-body system in Newtonian mechanics.

\item Level II.B -- {\it Theoretical knowledge in principle}:
The knowledge determined in principle by the fundamental deterministic laws,
even if in practice the system is too complicated to solve the fundamental
equations explicitly. For example, it may be a chaotic system with many relevant 
degrees of freedom.
 
\end{itemize}

\end{itemize}

In this paper we shall use all these levels of knowledge, depending on the context.
Thus, to avoid confusion, it is useful to have in mind which level of knowledge is used
when different probabilities are assigned to a given physical system.

It is also fair to note there is no sharp boundary between the different levels
of knowledge above. In fact, to obtain knowledge, one often combines knowledge
obtained from different levels. For instance, suppose that one wants to measure electric 
field $E$. To measure it, one actually measures (Level I)
the electric force $F$ on the charge $Q$, and then calculates (Level II)
the electric field by the formula $E=F/Q$. Essentially, this is why Einstein famously said 
that ``it is theory that decides what can be observed''.

 \subsection{What is special about time?}
\label{SEC2.1'}

In physics, one often takes for granted that time is a known variable, 
not subject to a probabilistic description. In particular, in QM one often
{\em requires} the pure state to have a unitary time evolution, because
otherwise the theory does not seem physically acceptable.
 
But why exactly is that so? What makes time so special, in a sense 
more certain than anything else?
Some may consider it to be intuitively obvious, others may find the question
too difficult, and yet others may believe that the question is philosophic rather 
than physical. In any case, we are not aware of any clear physical explanation available
in the literature. In this subsection we offer a simple physical justification
for such a view of time, based exclusively on classical physics. 

For that purpose, we consider a simple thought experiment. Consider a space-tourist traveling
in a spaceship, sitting in a compartment without windows or any other instruments
which could tell him about the world outside of the spaceship. The spaceship 
moves with non-relativistic velocities, having some complicated trajectory $X^i(t)$.
Ultimately (i.e. at Level II.B of knowledge) the trajectory $X^i(t)$ is determined by 
some deterministic laws, but at the practical level (Level II.A) 
the trajectory is too complicated for the tourist 
to calculate it. What can the tourist know about his position in space and time?

Since the knowledge is not available to him at Level II.A, he must resort to Level I,
that is, measurement. But since he has no windows or any other instruments
which could tell him about the outside world, he cannot determine where he is. In other 
words, he cannot measure $X^i$. Still, there is something which he {\em can} know
without any access to the outside world; he can determine time $t$ by measuring it
with his local clock. This provides a simple reason why time $t$ is better known
than the space position $X^i$.
(A nit-picking reader might object that the tourist can also measure his own acceleration,
from which he could reproduce the trajectory $X^i(t)$. But in general this is
not so simple. First, the change of velocity may, in part, be due to a free falling
around massive astrophysical objects, which the local accelerometer cannot detect.
Second, the spaceship engines may change the spaceship velocity very slowly, below the 
sensitivity threshold of a realistic accelerometer. Third, the calculation of the trajectory 
may be too complicated to perform in practice.)

Since the actual position $X^i$ of the tourist is not known to him, 
the best he can do is to assign some probability to any possible position $x^i$.
But since time is known to him, he will not assign probabilities to different
values of $t$. This means that his probabilistic description will be given
by some probability density $\rho({\bf x};t)$ satisfying
\begin{equation}\label{e2.4}
\int d^3x \, \rho({\bf x};t) =1 .  
\end{equation}
Essentially, this is why probability is conserved {\em in time}.

Now consider a variation of the thought experiment, with the only difference
that velocities of the spaceship may be relativistic. Now the trajectory
of the spaceship is some complicated function $X^{\mu}(\tau)$, where 
$\tau$ is the relativistic {\em proper time} measured by the clock in the spaceship.
Therefore, in the relativistic case it is the proper time $\tau$ that is known
to the tourist, and not the coordinate time $X^{0}$. The function 
$X^{0}(\tau)$ may be too complicated to calculate from first principles. 
The time $X^0$ can perhaps be measured by an external clock at rest 
with respect to the Lorentz
frame with coordinates $x^{\mu}$, but it cannot be measured by the tourist.
Therefore, at any given $\tau$, the tourist will assign some probabilities 
not only to the values of $x^i$, but also to the values of $x^0$.
In other words, his probabilistic description will be given
by some probability density $\rho(x;\tau)$ satisfying
\begin{equation}\label{e2.5}
\int d^4x \, \rho(x;\tau) =1 .  
\end{equation}
This is the simplest way to understand why, in relativity, one has to use
the {\em spacetime probability density conserved in proper time}
(Eq.~(\ref{e2.5})), and not
the space probability density conserved in coordinate time (Eq.~(\ref{e2.4})).

 \subsection{Equivariance: the simplest case}
\label{SEC2.3}

As noted in Introduction, 
the difference between space probability density and spacetime probability density
is related to a difference between space equivariance (\ref{e1}) and 
spacetime equivariance (\ref{e2}). Here we want to understand this difference
in the simplest possible case, by using Level II.B knowledge applied to a single 
classical particle.

Consider a non-relativistic particle with a known trajectory $X^i(t)$.
Since the trajectory is known, the probability density $\rho({\bf x};t)$ 
is a $\delta$-function 
\begin{equation}\label{e2.6}
 \rho({\bf x};t)=\delta^3({\bf x}-{\bf X}(t)) .
\end{equation}
This implies
\begin{equation}\label{e2.7}
 \frac{\partial\rho}{\partial t}=\frac{\partial\rho}{\partial X^i} \dot{X}^i(t).
\end{equation}
Here $\dot{X}^i(t)=dX^i(t)/dt$, while 
$\partial\rho/\partial X^i$ is actually $-\partial\rho/\partial x^i$, since
the right hand side of (\ref{e2.6}) is a function of ${\bf x}-{\bf X}$.
Introducing the notation $\partial_i=\partial/\partial x^i$ and noting
that $\dot{X}^i(t)$ depends only on $t$ so that $\partial_i\dot{X}^i=0$, 
we finally write (\ref{e2.7}) as 
\begin{equation}\label{e2.8}
 \frac{\partial\rho}{\partial t}+\partial_i(\rho \dot{X}^i) =0.
\end{equation}
This is nothing but the space-equivariance equation for the probability density (\ref{e2.6}).

Now consider a relativistic particle with a known trajectory  $X^{\mu}(\tau)$.
Instead of (\ref{e2.6}), now we have a spacetime probability 
density
\begin{equation}\label{e2.9}
 \rho(x;\tau)=\delta^4(x-X(\tau)) .
\end{equation}
Thus, in a completely analogous way, instead of (\ref{e2.8}) we obtain
\begin{equation}\label{e2.10}
 \frac{\partial\rho}{\partial\tau}+\partial_{\mu}(\rho \dot{X}^{\mu}) =0,
\end{equation}
where $\dot{X}^{\mu}(\tau)=dX^{\mu}(\tau)/d\tau$. Eq.~(\ref{e2.10}) is the 
simplest example of a spacetime-equivariance equation.

To see the relation between space equivariance and spacetime equivariance,
it is instructive to see how the former can be derived from the latter.
For that purpose we first write (\ref{e2.9}) as
\begin{equation}\label{e2.11}
 \rho(x;\tau)=\delta(x^0-X^0(\tau)) \, \rho^{(3)}({\bf x};\tau),
\end{equation}
where 
\begin{equation}\label{e2.12}
\rho^{(3)}({\bf x};\tau)=\delta^3({\bf x}-{\bf X}(\tau)) .
\end{equation}
Then, to eliminate $x^0$ from (\ref{e2.10}), we integrate (\ref{e2.10}) over $\int dx^0$.
Since
\begin{equation}\label{e2.13}
 \int dx^0 \partial_{0}(\rho \dot{X}^{0}) =0 , \;\;\;\;
 \int dx^0 \rho(x;\tau) = \rho^{(3)}({\bf x};\tau) ,
\end{equation}
we see that the integration of (\ref{e2.10}) leads to
\begin{equation}\label{e2.14}
 \frac{\partial\rho^{(3)}}{\partial\tau}+\partial_i(\rho^{(3)} \dot{X}^{i}) =0 .
\end{equation}
Eqs.~(\ref{e2.12}) and (\ref{e2.14}) are very similar to (\ref{e2.6}) and (\ref{e2.8}),
respectively. Thus, (\ref{e2.14}) represents the relativistic  
space-equivariance equation, with the relativistic proper time $\tau$ instead of the
non-relativistic time $t$ in (\ref{e2.8}). Even though (\ref{e2.14}) is relativistic,
note, however, that it is not manifestly covariant, but valid in one fixed set 
of coordinates $x^0$, $x^i$. If one transforms it to another set of coordinates,
then (\ref{e2.14}) transforms into an equation having a different form.

Finally, note that in the non-relativistic limit the proper time $\tau$ becomes 
identical to the non-relativistic time $t$. Thus, in the non-relativistic
limit the space-equivariance equation (\ref{e2.14}) becomes identical to (\ref{e2.8}).

\section{Time and probability in non-relativistic mechanics} 
\label{SEC3}

 \subsection{Hamilton-Jacobi equation}
\label{SEC3.1}

Consider $n$ non-relativistic and mutually non-interacting particles, each described
by its own Hamiltonian
\begin{equation}\label{e3.1}
 H_a=\frac{{\bf p}_a^2}{2m_a}+U_a({\bf x}_a) ,
\end{equation}
where $a=1,\ldots ,n$ are the particle labels, ${\bf p}_a$ is the momentum of the particle
with position ${\bf x}_a$, and $U_a({\bf x}_a)$ is the potential for the particle labeled by
$a$. Since the $n$ systems do not interact with each other, their 
internal clocks cannot be synchronized. Hence it is justified to use 
a different time variable $t_a$ for each particle trajectory ${\bf X}_a(t_a)$.
Therefore the system can be described by $n$ Hamilton-Jacobi equations
\begin{equation}\label{e3.2}
 \frac{(\partial_{ai}S_a)(\partial_{ai}S_a)}{2m_a}+U_a({\bf x}_a)=
-\frac{\partial S_a}{\partial t_a} ,
\end{equation}
one for each $a$. Here $S_a({\bf x}_a,t_a)$ is the principal function,
$\partial_{ai}=\partial/\partial x_a^i$, and the summation over repeated index 
$i$ is understood. Introducing the velocity field
\begin{equation}\label{e3.3}
 v_a^i({\bf x}_a,t_a)=\frac{\partial_{ai}S_a({\bf x}_a,t_a)}{m_a} ,
\end{equation}
the particle trajectories are determined by
\begin{equation}\label{e3.4}
 \frac{dX_a^i(t_a)}{dt_a}=v_a^i({\bf X}_a(t_a),t_a) .
\end{equation}

The above represents the Hamilton-Jacobi formalism in the many-time formulation.
The usual single-time formulation may be recovered in the following way.
One performs the sum $\sum_a$ in (\ref{e3.2}) to obtain
\begin{equation}\label{e3.5}
 \sum_{a=1}^{n} \frac{(\partial_{ai}S)(\partial_{ai}S)}{2m_a}+
U({\bf x}_1,\ldots,{\bf x}_n)=
-\sum_{a=1}^{n}\frac{\partial S}{\partial t_a} ,
\end{equation}
where
\begin{equation}\label{e3.6}
  U({\bf x}_1,\ldots,{\bf x}_n)=\sum_{a=1}^{n} U_a({\bf x}_a),
\end{equation}
\begin{equation}\label{e3.7}
 S({\bf x}_1,t_1,\ldots,{\bf x}_n,t_n) =\sum_{a=1}^{n} S_a({\bf x}_a,t_a) .
\end{equation}

The next step is to consider the
special case in which all time variables $t_a$ have the same value $t$.
For that purpose we use the following identity. Let $f(t_1,\ldots,t_n)$
be an arbitrary function, and let
\begin{equation}\label{e3.8}
 f(t)\equiv f(t_1,\ldots,t_n)|_{t_1=\cdots=t_n=t} .
\end{equation}
Then
\begin{equation}\label{identity}
\left. 
\sum_{a=1}^{n}\frac{\partial f(t_1,\ldots,t_n)}{\partial t_a} \right|_{t_1=\cdots=t_n=t}
=\frac{\partial f(t)}{\partial t} .
\end{equation}
In this paper, we shall use the many-time identity (\ref{identity}) 
many times.\footnote{Pun intended.}

Applying the identity (\ref{identity}) to (\ref{e3.5}), we obtain the
standard single-time Hamilton-Jacobi equation
\begin{equation}\label{e3.9}
 \sum_{a=1}^{n} \frac{(\partial_{ai}S)(\partial_{ai}S)}{2m_a}+
U({\bf x}_1,\ldots,{\bf x}_n)= -\frac{\partial S}{\partial t} 
\end{equation}
for the single-time principal function 
\begin{equation}\label{e3.10}
 S({\bf x}_1,\ldots,{\bf x}_n,t)=S({\bf x}_1,t_1,\ldots,{\bf x}_n,t_n) |_{t_1=\cdots=t_n=t} .
\end{equation}
Similarly, (\ref{e3.3}) and (\ref{e3.4}) reduce to the usual single-time equations       
\begin{equation}\label{e3.11}
 v_a^i({\bf x}_a,t)=\frac{\partial_{ai}S({\bf x}_1,\ldots,{\bf x}_n,t)}{m_a} ,
\end{equation}
\begin{equation}\label{e3.12}
 \frac{dX_a^i(t)}{dt}=v_a^i({\bf X}_a(t),t) .
\end{equation}

Note that in (\ref{e3.12}) the velocity of particle $a$ does not depend
on the positions of other particles. This is a manifestation of the fact that
classical mechanics is local, which can be traced back to the locality
of the classical potential (\ref{e3.6}). In particular, even though all trajectories
can be parameterized by the same time variable $t$ in (\ref{e3.12}),
it does not mean that the theory is not local. 

 \subsection{Space equivariance}
\label{SEC3.2}
     
Now, instead of a single trajectory ${\bf X}_a(t_a)$, consider a statistical ensemble 
of such trajectories, with velocities determined by the velocity field (\ref{e3.3}). 
In the ensemble, let $\rho_a({\bf x}_a;t_a)$ be the probability density
that the particle has position ${\bf x}_a$ at time $t_a$. 
To derive the space-equivariance equation, consider a 3-dimensional region 
of space ${\cal R}_a^{(3)}$ with the 2-dimensional boundary $\partial{\cal R}_a^{(3)}$. 
The number of ensemble particles escaping from ${\cal R}_a^{(3)}$ per unit time is
\begin{equation}\label{e3.13}
 \int_{\partial{\cal R}_a^{(3)}} dA_{ai} \, \rho_a v_a^i = 
\int_{{\cal R}_a^{(3)}} d^3x_a \, \partial_{ai}(\rho_a v_a^i) , 
\end{equation}
where $dA_{ai}$ is the surface element on the boundary and the Gauss theorem has been used.
But the particles in the ensemble are neither created nor destroyed.
The number of particles (per unit time) that disappeared in ${\cal R}_a^{(3)}$ 
must be equal to the number of particles (per unit time) that escaped from ${\cal R}_a^{(3)}$,
i.e.,
\begin{equation}\label{e3.14}
- \int_{{\cal R}_a^{(3)}} d^3x_a \, \frac{\partial\rho_a}{\partial t_a}
=  \int_{{\cal R}_a^{(3)}} d^3x_a \, \partial_{ai}(\rho_a v_a^i) .
\end{equation}
But this must be valid for {\em any} ${\cal R}_a^{(3)}$, which leads to the local
equation 
\begin{equation}\label{e3.15}
\frac{\partial\rho_a}{\partial t_a} + \partial_{ai}(\rho_a v_a^i) =0.
\end{equation}
Eq.~(\ref{e3.15}) is referred to as the {\em space-equivariance equation}.
 
A note on terminology is in order. In the literature, Eq.~(\ref{e3.15}) is also called
``continuity equation'' or ``local conservation equation''. The name
``continuity equation'' is perhaps better suited in the context of continuous fluids,
while the name ``local conservation equation'' is perhaps better suited
to describe the conservation of quantities such as charge, the density $\rho$ of which
does not need to be positive. Thus, to avoid confusion, in this paper we use the name
``equivariance equation'' when we refer to a property of probability density $\rho$ 
associated with a statistical ensemble.

Here $\rho_a({\bf x}_a;t_a)$ is the probability density for a {\em single} particle.
But the label $a$ reminds us that we have $n$ particles, each described
by a separate probability density $\rho_a$. 
We assume that the particles do not interact with each other
(see (\ref{e3.6})), which means that the particles cannot be correlated.
Therefore the joint probability density for all particles is
\begin{equation}\label{e3.16}
 \rho({\bf x}_1,\ldots,{\bf x}_n;t_1,\ldots,t_n)=
\rho_1({\bf x}_1;t_1) \cdots \rho_n({\bf x}_n;t_n).
\end{equation}
This is the joint probability density that the first particle will have the position
${\bf x}_1$ at time $t_1$, second particle the position
${\bf x}_2$ at time $t_2$, etc.
Thus (\ref{e3.15}) can also be written as a many-time equivariance equation for the
joint probability density
\begin{equation}\label{e3.17}
\frac{\partial\rho}{\partial t_a} + \partial_{ai}(\rho v_a^i) =0.
\end{equation}
 
Eq.~(\ref{e3.17}) can also be written in the more familiar single-time form,
by considering the special case $t_1=\cdots=t_n=t$. In this way, by summing
(\ref{e3.17}) over $a$ and using the identity (\ref{identity}), we obtain
\begin{equation}\label{e3.18}
\frac{\partial\rho}{\partial t} + \sum_{a=1}^{n} \partial_{ai}(\rho v_a^i) =0,
\end{equation}
where
\begin{equation}\label{e3.19}
 \rho({\bf x}_1,\ldots,{\bf x}_n;t)=
\rho({\bf x}_1,\ldots,{\bf x}_n;t_1,\ldots,t_n) |_{t_1=\cdots=t_n=t} .
\end{equation}

 \subsection{Schr\"odinger equation}
\label{SEC3.3}
     
Let $\psi({\bf x}_1,t_1,\ldots,{\bf x}_n,t_n)$ be the many-time wave function
associated with a system described by a quantum variant of the Hamiltonians (\ref{e3.1}).
It satisfies $n$ Schr\"odinger equations
\begin{equation}\label{e3.20}
 \hat{H}_a \psi = i\frac{\partial\psi}{\partial t_a} , 
\end{equation}
where we use units $\hbar=1$. By writing $\psi$ in the polar form 
\begin{equation}\label{e3.21}
  \psi({\bf x}_1,t_1,\ldots,{\bf x}_n,t_n)
=R({\bf x}_1,t_1,\ldots,{\bf x}_n,t_n) e^{iS({\bf x}_1,t_1,\ldots,{\bf x}_n,t_n)} ,
\end{equation}
the Schr\"odinger equations can be written as
\begin{equation}\label{e3.22}
\frac{(\partial_{ai}S)(\partial_{ai}S)}{2m_a}+U_a({\bf x}_a)+Q_a=
-\frac{\partial S}{\partial t_a} ,
\end{equation}
\begin{equation}\label{e3.23}
\frac{\partial R^2}{\partial t_a} + \partial_{ai}(R^2 v_a^i) =0 , 
\end{equation}
where
\begin{equation}\label{e3.24}
 Q_a=-\frac{1}{2m_a} \frac{\partial_{ai}\partial_{ai}R}{R} ,
\end{equation}
\begin{equation}\label{e3.25}
 v_a^i=\frac{\partial_{ai}S}{m_a} .
\end{equation}

Note that, in general, $\psi({\bf x}_1,t_1,\ldots,{\bf x}_n,t_n)$ does not need
to be a product of the form $\prod_a \psi_a({\bf x}_a,t_a)$. 
Therefore, the joint probability density
\begin{equation}\label{e3.26}
 \rho({\bf x}_1,\ldots,{\bf x}_n;t_1,\ldots,t_n)=
|\psi({\bf x}_1,t_1,\ldots,{\bf x}_n,t_n)|^2 
\end{equation}
does not need to have the product form (\ref{e3.16}), which, of course,
is a manifestation of quantum nonlocality. Likewise,
$R$ does not need to be a product $\prod_a R_a({\bf x}_a,t_a)$ and $S$ does 
not need to be a sum of the form (\ref{e3.7}). 
As a consequence, (\ref{e3.24}) and (\ref{e3.25}) do not need to be local quantities
of the form $Q_a({\bf x}_a,t_a)$ and $v_a^i({\bf x}_a,t_a)$, respectively.
This implies that, in general, (\ref{e3.22}) {\em cannot} be interpreted as a
quantum Hamilton-Jacobi equation for the particle $a$, and (\ref{e3.25})
{\em cannot} be used to calculate a many-time Bohmian velocity $dX_a^i(t_a)/dt_a$.  

Instead, to obtain a consistent interpretation in terms of Bohmian particle
trajectories, one needs to reintroduce the more familiar single-time formulation.
By summing (\ref{e3.20}) over $a$ and using the identity (\ref{identity}), one obtains
the single-time Schr\"odinger equation
\begin{equation}\label{e3.27}
 \hat{H}\psi = i\frac{\partial\psi}{\partial t} , 
\end{equation}
where $\hat{H}=\sum_a \hat{H}_a$. 
In a similar way, (\ref{e3.22}) and (\ref{e3.23}) become 
\begin{equation}\label{e3.28}
\sum_{a=1}^n \frac{(\partial_{ai}S)(\partial_{ai}S)}{2m_a}+U+Q=
-\frac{\partial S}{\partial t} ,
\end{equation}
\begin{equation}\label{e3.29}
\frac{\partial R^2}{\partial t} + \sum_{a=1}^n\partial_{ai}(R^2 v_a^i) =0 , 
\end{equation}
where $U$ is given by (\ref{e3.6}) and $Q=\sum_a Q_a$.  
Now (\ref{e3.29}) can be interpreted as the space-equivariance equation, provided that
particles have the trajectories $X_a^i(t)$ (all parameterized by the same time parameter $t$),
with the usual Bohmian velocities
\begin{equation}\label{e3.30}
\frac{dX_a^i}{dt}=v_a^i .
\end{equation}
Unlike the velocity in the classical case (\ref{e3.12}),
here the quantum velocity $v_a^i({\bf x}_1,\ldots,{\bf x}_n,t)$ is a nonlocal function 
depending on all particle positions. 

 \subsection{Newton time as non-relativistic proper time}
\label{SEC3.4}

     \subsubsection{Non-relativistic proper time in classical mechanics}
\label{SEC3.4.1}

Eq.~(\ref{e3.4}) implies
\begin{equation}\label{e3.31}
 \frac{dX_a^i dX_a^i}{dt_a^2}=v_a^i v_a^i ,
\end{equation}
where the summation over $i$ is understood. It is suggestive to rewrite it as
\begin{equation}\label{e3.32}
 dt_a^2 = \frac{1}{v_a^j v_a^j} dX_a^i dX_a^i ,
\end{equation}
because in this form it looks similar to (\ref{e5}). The similarity between 
(\ref{e3.32}) and (\ref{e5}) suggests the interpretation of $t_a$ as 
{\em non-relativistic proper time}. Here the word ``proper'' refers to 
the fact $dt_a^2$ is proportional to $dX_a^i dX_a^i$, implying that,
similarly to the familiar relativistic proper time $\tau$,
$t_a$ is defined along the {\em trajectory} of the particle $a$.
In fact, (\ref{e3.32}) is even more similar
to (\ref{e6}), which is obvious if we write (\ref{e3.32}) as
\begin{equation}\label{e3.33}
 dt_a^2 = f_a({\bf x}_a,t_a) \, dX_a^i dX_a^i ,
\end{equation}
where $f_a=1/v_a^j v_a^j$.

Even though each particle has its own non-relativistic proper time $t_a$,
one can also consider the case in which all variables $t_a$ have the same value $t$.
In this special case one can write (\ref{e3.32}) and  (\ref{e3.33}) as
\begin{equation}\label{e3.34}
 dt^2 = \frac{1}{v_a^j v_a^j} dX_a^i dX_a^i = f_a({\bf x}_a,t) \, dX_a^i dX_a^i ,
\end{equation}
which can also be obtained more directly from (\ref{e3.12}).
Even though in (\ref{e3.34}) one uses the same time parameter $t$ for all the particles, 
this non-relativistic proper time is local, which is encoded in the fact that
$f_a({\bf x}_a,t)$ does not depend on positions of other particles.

     \subsubsection{Non-relativistic proper time in Bohmian mechanics}
\label{SEC3.4.2}

Non-relativistic proper time in Bohmian mechanics can be introduced in a way
completely analogous to that in classical mechanics. 
From (\ref{e3.30}) one obtains 
\begin{equation}\label{e3.35}
 dt^2 = \frac{1}{v_a^j v_a^j} dX_a^i dX_a^i = 
f_a({\bf x}_1,\ldots,{\bf x}_n,t) \, dX_a^i dX_a^i .
\end{equation}
This has the same form as the classical non-relativistic proper time 
(\ref{e3.34}), with one crucial difference. 
In the quantum case the non-relativistic proper time is not local, which is encoded
in the fact that
$f_a({\bf x}_1,\ldots,{\bf x}_n,t)$ depends on the positions of all particles.

Note also the obvious similarity between (\ref{e3.35}) and (\ref{e6}). In particular, 
both expressions for proper times are nonlocal.

\section{Time and probability in classical relativistic mechanics}
\label{SEC4}

 \subsection{Relativistic Hamilton-Jacobi equation}
\label{SEC4.1}

Consider $n$ free classical relativistic particles. Each can be described by a
relativistic Hamilton-Jacobi equation (see e.g. \cite{nikBOOK12}) 
\begin{equation}\label{e3.36}
 (\partial_{a\mu}S_a)(\partial_{a}^{\mu}S_a)-m_a^2=0 , 
\end{equation}
where the summation over repeated spacetime index ${\mu}$ is understood,
$\partial_{a\mu} = \partial/\partial x_{a}^{\mu}$, 
$\partial_{a}^{\mu}=\eta^{\mu\nu}\partial_{a\nu}$, and $S_a(x_a)$ is the relativistic
principal function. The corresponding velocity field is
\begin{equation}\label{e3.37}
 V_{a}^{\mu}(x_a)=-\frac{\partial_{a}^{\mu}S_a(x_a)}{m_a} ,
\end{equation}
and the corresponding particle trajectories are given by
\begin{equation}\label{e3.38}
  \frac{dX_a^{\mu}(\tau_a)}{d\tau_a}=V_{a}^{\mu}(X_a(\tau_a)) .
\end{equation}

Each particle trajectory is parameterized by its own scalar parameter $\tau_a$.
(In the next subsection we shall see that $\tau_a$ is in fact
the standard classical relativistic proper time, but for the moment we can ignore it.)
Nevertheless, analogously to the non-relativistic case in Sec.~\ref{SEC3}, 
we can also study a special case in which all scalar parameters $\tau_a$  
have the same value $\tau$. Eqs.~(\ref{e3.36}) and (\ref{e3.37}) have not
any explicit dependence on $\tau_a$, so the only equation that gets modified is 
(\ref{e3.38}), which becomes
\begin{equation}\label{e3.39}
  \frac{dX_a^{\mu}(\tau)}{d\tau}=V_{a}^{\mu}(X_a(\tau)) .
\end{equation} 
Even though we use the same parameter $\tau$ for all the particles, 
the dynamics is still local, owing to the fact that the right-hand side
of (\ref{e3.39}) does not depend on the positions of other particles.

Finally, we can sum (\ref{e3.36}) over $a$ and introduce the function
\begin{equation}\label{e3.40}
 S(x_1,\ldots,x_n) = \sum_{a=1}^n S_a(x_a) .
\end{equation}
In this way (\ref{e3.36}) leads to
\begin{equation}\label{e3.41}
 \sum_{a=1}^n [(\partial_{a\mu}S)(\partial_{a}^{\mu}S)+m_a^2]=0 , 
\end{equation}
and (\ref{e3.37}) can be written as
\begin{equation}\label{e3.42}
 V_{a}^{\mu}(x_a)=-\frac{\partial_{a}^{\mu}S(x_1,\ldots,x_n)}{m_a} .
\end{equation}

 \subsection{Relativistic proper time}
\label{SEC4.2}

Eq.~(\ref{e3.38}) implies 
\begin{equation}\label{e3.43}
\frac{dX_{a\mu}dX_a^{\mu}}{d\tau_a^2}=V_{a\mu}V_{a}^{\mu} ,
\end{equation}
which can be written in the form analogous to (\ref{e3.32})
\begin{equation}\label{e3.44}
 d\tau_a^2 = \frac{1}{V_{a\nu}V_{a}^{\nu}} dX_{a\mu}dX_a^{\mu} .
\end{equation}
However, using (\ref{e3.37}), we see that the Hamilton-Jacobi equation (\ref{e3.36}) 
can be written as
\begin{equation}\label{e3.45}
 V_{a\mu}V_{a}^{\mu}=1.
\end{equation}
In this way (\ref{e3.44}) simplifies to
\begin{equation}\label{e3.46}
 d\tau_a^2 =  dX_{a\mu}dX_a^{\mu} = \eta_{\mu\nu} dX_a^{\mu} dX_a^{\nu},
\end{equation}
which shows that $\tau_a$ is nothing but the standard classical relativistic 
proper time along the particle trajectory.

Alternatively, if we use (\ref{e3.39}), we obtain
\begin{equation}\label{e3.47.1}
 d\tau^2 =\frac{1}{V_{a\nu}V_{a}^{\nu}} dX_{a\mu}dX_a^{\mu} ,
\end{equation}
reflecting again the fact that one can use the same value of proper time for all
the particles.
Using (\ref{e3.45}), we see that this can also be written in the form of (\ref{e3.46}), i.e.
\begin{equation}\label{e3.47}
 d\tau^2 =  \eta_{\mu\nu} dX_a^{\mu} dX_a^{\nu} .
\end{equation}

 \subsection{Spacetime equivariance} 
\label{SEC4.3}

Consider a statistical ensemble of relativistic particles, with velocities determined 
by the velocity field (\ref{e3.37}). In the ensemble, let 
$\rho_a(x_a;\tau_a)$ be the probability density
that the particle has the spacetime position $x_a$ at time $\tau_a$.
By considering a 4-dimensional region 
of spacetime ${\cal R}_a^{(4)}$ and performing a procedure completely analogous
to that in Sec.~\ref{SEC3.2}, one obtains the
spacetime-equivariance equation
\begin{equation}\label{e3.48}
\frac{\partial\rho_a}{\partial \tau_a} + \partial_{a\mu}(\rho_a V_a^{\mu}) =0 ,
\end{equation}
which is the relativistic generalization of the space-equivariance equation (\ref{e3.15}).
Similarly, by introducing 
\begin{equation}\label{e3.49}
 \rho(x_1,\ldots,x_n;\tau_1,\ldots,\tau_n)=
\rho_1(x_1;\tau_1) \cdots \rho_n(x_n;\tau_n) ,
\end{equation}
one obtains the relativistic generalization of (\ref{e3.17})
\begin{equation}\label{e3.50}
\frac{\partial\rho}{\partial \tau_a} + \partial_{a\mu}(\rho V_a^{\mu}) =0.
\end{equation}

Finally, by considering the case $\tau_1=\cdots=\tau_n=\tau$ and summing over $a$,
one obtains the relativistic generalizations of (\ref{e3.18}) and (\ref{e3.19})
\begin{equation}\label{e3.51}
\frac{\partial\rho}{\partial \tau} + \sum_{a=1}^{n} \partial_{a\mu}(\rho V_a^{\mu}) =0,
\end{equation}
\begin{equation}\label{e3.52}
 \rho(x_1,\ldots,x_n;\tau)=
\rho(x_1,\ldots,x_n;\tau_1,\ldots,\tau_n) |_{\tau_1=\cdots=\tau_n=\tau} .
\end{equation}

 \subsection{Space equivariance from spacetime equivariance}
\label{SEC4.4}

     \subsubsection{Does space probability imply space equivariance?}
\label{SEC4.4.1}

The simplest way to get space probability density from spacetime probability density
is to consider conditional probability. For simplicity, consider the case 
of a single particle. Suppose that the value of $X^0$ is somehow determined
for some fixed value of $\tau={\cal T}$. The values of $X^i$ are still unknown, so 
knowledge about the system at proper time ${\cal T}$
is described by the conditional probability density
\begin{equation}\label{ep1}
\rho_{\rm cond}({\bf x};X^0,{\cal T})=
\frac{\rho({\bf x},X^0;{\cal T})}{N(X^0)} ,
\end{equation} 
where
\begin{equation}
N(X^0)=\int d^3x \, \rho({\bf x},X^0;{\cal T})
\end{equation}
does not depend on ${\cal T}$ due to (\ref{e3.51}). 
Does the space probability density (\ref{ep1}) satisfy a space equivariance equation?

To make the question meaningful, one must define a time-evolution of the 
conditional probability. How to do that? 

One possibility is to keep $\tau={\cal T}$ fixed
and vary $X^0$. However, determination of $X^0$ effectively reduces
the initial statistical ensemble to a particular {\em sub-ensemble}, in which 
all members have the same value of $X^0$. To {\em vary} $X^0$ means to pick out
{\em another} sub-ensemble with a {\em different} value of $X^0$. On the other hand,
the equivariance equation refers to an evolution of a {\em fixed} statistical
ensemble. Therefore, in general, such an evolution of conditional 
probability is not expected to obey an equivariance equation.

What if one keeps $X^0$ fixed but varies $\tau$ instead? That would mean that 
for each value of $\tau$ one must perform an independent determination of $x^0$ 
and discard the result of determination unless the value of $x^0$
is equal to the fixed value $X^0$. But this again 
means that for each value of $\tau$ one must pick out a different 
sub-ensemble (the one which for {\em that} value of $\tau$ attains the value
$X^0$), implying that an equivariance equation is not to be expected.  

Therefore, in general, the simplest way to get space probability density
from spacetime probability density does not lead to a space-equivariance equation.
Nevertheless, there are two special cases in which spacetime equivariance 
does imply space equivariance. We discuss these two cases in the following Secs.~\ref{SEC4.4.2} and \ref{SEC4.4.3}.

     \subsubsection{Ensemble with fixed $X_a^0(\tau)$}
\label{SEC4.4.2}

Let us start with a statistical ensemble satisfying the spacetime-equivariance 
equation (\ref{e3.51}). In addition,
suppose that in the ensemble $X_a^0(\tau)$ is fixed, i.e., that all particles labeled 
by $a=1$ have the same $X_1^0(\tau)$, all particles labeled 
by $a=2$ have the same $X_2^0(\tau)$, etc. This means that $\rho$ has the form
\begin{equation}\label{e2.11n}
 \rho(x_1,\ldots,x_n;\tau)=\delta(x_1^0-X_1^0(\tau)) \cdots \delta(x_n^0-X_n^0(\tau))
\, \rho^{(3)}({\bf x}_1,\ldots,{\bf x}_n;\tau),
\end{equation}
which is a generalization of (\ref{e2.11}).
Indeed, from (\ref{e2.11n}) we see that
\begin{equation}\label{e2.13n}
 \int dx_1^0  \cdots \int dx_n^0 \,  \rho(x_1,\ldots,x_n;\tau)= 
\rho^{(3)}({\bf x}_1,\ldots,{\bf x}_n;\tau) ,
\end{equation}
which generalizes the second equation in (\ref{e2.13}). 
More generally, for an arbitrary function 
$F({\bf x}_1,x^0_1,\ldots,{\bf x}_n,x^0_n,\tau)$
we have
\begin{eqnarray}
& \displaystyle\int dx_1^0 \,  \delta(x_1^0-X_1^0(\tau)) 
\cdots \int dx_n^0 \, \delta(x_n^0-X_n^0(\tau)) \,
F({\bf x}_1,x^0_1,\ldots,{\bf x}_n,x^0_n,\tau) &
\nonumber \\
& = F({\bf x}_1,X^0_1(\tau),\ldots,{\bf x}_n,X^0_n(\tau),\tau) \equiv
\tilde{F}({\bf x}_1,\ldots,{\bf x}_n,\tau) . &
\end{eqnarray}
In this way, by integrating  
(\ref{e3.51}) over $\int dx_1^0  \cdots \int dx_n^0$ we get
\begin{equation}\label{e3.51_3}
\frac{\partial\rho^{(3)}}{\partial \tau} + 
\sum_{a=1}^{n} \partial_{ai}(\rho^{(3)} \tilde{V}_a^{i}) =0,
\end{equation}
which is a generalization of (\ref{e2.14}). 
But from (\ref{e2.11n}) we see that $\rho^{(3)}$ is the space probability density
(for that purpose, see also (\ref{e2.1})-(\ref{e2.3})).
This implies that (\ref{e3.51_3}) is indeed the space-equivariance equation
for the space probability density $\rho^{(3)}$. 

As a special case, consider the non-relativistic limit. In this limit 
$d\tau=dt$, so (\ref{e3.51_3}) can be written as
\begin{equation}
 \frac{\partial\rho^{(3)}}{\partial t} + 
\sum_{a=1}^{n} \partial_{ai}(\rho^{(3)} \tilde{V}_a^{i}) =0 .
\end{equation}

     \subsubsection{Stationary ensemble with constant $V^0_a$}
\label{SEC4.4.3}

As a substantially different case in which spacetime equivariance leads to space equivariance,
assume that the ensemble is stationary, i.e., that 
\begin{equation}
 \frac{\partial\rho_a}{\partial \tau_a}=0. 
\end{equation}
This means that (\ref{e3.48}) reduces to  
\begin{equation}\label{e3.48_st1}
\partial_{a\mu}(\rho_a V_a^{\mu}) =0 .
\end{equation}
Furthermore, assume that $V^0_a$ is constant, i.e., does not depend on $x_a$. This means that
(\ref{e3.48_st1}) can be written as
\begin{equation}\label{e3.48_st2}
V_a^{0} \partial_{a0} \rho_a   + \partial_{ai}(\rho_a V_a^{i}) =0 .
\end{equation}
By dividing it with the constant $V^0_a$, we get 
\begin{equation}\label{e3.48_st3}
\partial_{a0} \rho_a   + \partial_{ai}(\rho_a v_a^{i}) =0 ,
\end{equation} 
where
\begin{equation}\label{v3}
 v_a^{i}=\frac{V_a^{i}}{V_a^{0}} .
\end{equation}
Particle trajectories satisfy 
\begin{equation}
 \frac{dX_a^i}{dX_a^0}=\frac{dX_a^i/d\tau_a}{dX_a^0/d\tau_a}=\frac{V_a^{i}}{V_a^{0}}=v_a^{i},
\end{equation}
where in the last equality we used (\ref{v3}). This shows that $v_a^{i}$ in 
(\ref{e3.48_st3}) is the ``Newton'' velocity, i.e., velocity defined as a derivative 
with respect to coordinate time $X_a^0$ (not the proper time $\tau_a$).

Since $dX_a^0/d\tau_a=V_a^{0}$ is constant, the solution of this equation is simple:
\begin{equation}\label{simple}
 X_a^0(\tau_a)=Y_a^0+V_a^0 \tau_a,
\end{equation}
where $Y_a^0$ is an integration constant representing the initial condition.
The simplicity of (\ref{simple}) means that someone (e.g. the tourist from
Sec.~\ref{SEC2.1'}) who knows the initial
condition $Y_a^0$ can easily know $X_a^0(\tau_a)$ for each $\tau_a$, not only
in principle, but even in practice (i.e. at Level II.A in Sec.~\ref{SEC2.1}). 
Consequently, $x_a^0$ can be treated as a known variable, so the relevant 
probability density is not $\rho_a(x_a)$, but the conditional probability
\begin{equation}
\rho_a^{\rm cond}({\bf x}_a;x_a^0)=\frac{\rho_a({\bf x}_a,x_a^0)}{N_a} ,  
\end{equation}
where
\begin{equation}
 N_a=\int d^3x_a \, \rho_a({\bf x}_a,x_a^0)
\end{equation}
does not depend on $x_a^0$ due to (\ref{e3.48_st3}). Hence, one can divide
(\ref{e3.48_st3}) by $N_a$ to get
\begin{equation}\label{e3.48_st4}
\frac{\partial\rho_a^{\rm cond}}{\partial x_a^0}   
+ \partial_{ai}(\rho_a^{\rm cond} v_a^{i}) =0 ,
\end{equation}
which is the space-equivariance equation for the conditional probability 
$\rho_a^{\rm cond}$.

Finally, recalling ({\ref{e3.49}}), one can multiply (\ref{e3.48_st4})
with $\prod_{a'\neq a} \rho_{a'}^{\rm cond}$ to get 
\begin{equation}\label{e3.48_st5}
\frac{\partial\rho_{\rm cond}}{\partial x_a^0}   
+ \partial_{ai}(\rho_{\rm cond} v_a^{i}) =0 ,
\end{equation}  
where $ \rho_{\rm cond}=\prod_a \rho_a^{\rm cond}$. 
Putting $x_1^0=\cdots=x_n^0=t$, summing over $a$ and using (\ref{identity}), 
one gets the single-time space-equivariance equation
\begin{equation}\label{e3.48_st6}
\frac{\partial\rho_{\rm cond}}{\partial t}   
+ \sum_{a=1}^n \partial_{ai}(\rho_{\rm cond} v_a^{i}) =0 .
\end{equation}

The derivation above is based on classical relativistic mechanics, 
but in Sec.~\ref{SEC5.5} we shall see that, under certain conditions,
in relativistic Bohmian mechanics
a space-equivariance equation can be obtained in a very similar way.

\section{Time and probability in relativistic Bohmian mechanics}
\label{SEC5}

 \subsection{Klein-Gordon equation}
\label{SEC5.1}

The relativistic wave function $\psi(x_1,\ldots,x_n)$ 
for $n$ free relativistic particles without spin satisfies $n$ Klein-Gordon equations
\begin{equation}\label{e5.1}
 [\partial_{a\mu}\partial_{a}^{\mu}+m_a^2]\psi=0 . 
\end{equation}
It implies that the currents $j_{a\mu}(x_1,\ldots,x_n)$ defined as
\begin{equation}\label{e5.2}
j_{a\mu} = \frac{i}{2m_a} \, \psi^* \stackrel{\leftrightarrow\;\;}{\partial_{a\mu}} \psi 
\end{equation}
with
\begin{equation}\label{antisimder}
A\!\stackrel{\leftrightarrow\;}{\partial_{\mu}}\!B \equiv A (\partial_{\mu} B)-(\partial_{\mu} A) B ,
\end{equation}
are conserved
\begin{equation}\label{e5.3}
 \partial_{a\mu} j_a^{\mu}=0.
\end{equation}

By writing $\psi$ in the polar form 
\begin{equation}\label{e5.4}
  \psi(x_1,\ldots,x_n) =R(x_1,\ldots,x_n) e^{iS(x_1,\ldots,x_n)} ,
\end{equation}
the Klein-Gordon equations can be written as 
\begin{equation}\label{e5.5}
 (\partial_{a\mu}S)(\partial_{a}^{\mu}S)-m_a^2-2m_aQ_a=0 , 
\end{equation}
\begin{equation}\label{e5.6}
 \partial_{a\mu}(R^2 V_a^{\mu})=0 ,
\end{equation}
where
\begin{equation}\label{e5.7}
 Q_a=\frac{1}{2m_a}\frac{\partial_{a\mu}\partial_{a}^{\mu}R}{R} ,
\end{equation}
\begin{equation}\label{e5.8}
 V_{a}^{\mu}=-\frac{\partial_{a}^{\mu}S}{m_a} .
\end{equation}
Clearly, (\ref{e5.8}) is analogous to (\ref{e3.42}), with the difference that in (\ref{e5.8})
$V_{a}^{\mu}(x_1,\ldots,x_n)$ does not need to be a local function.
From (\ref{e5.2}) and (\ref{e5.8}) one can also see that
\begin{equation}\label{e5.9} 
j_a^{\mu}=R^2 V_a^{\mu} ,
\end{equation}
so (\ref{e5.8}) is equivalent to
\begin{equation}\label{e5.10}
 V_{a}^{\mu}=\frac{j_a^{\mu}}{\psi^*\psi} .
\end{equation}

The most general solution of (\ref{e5.1}) is an arbitrary superposition of plane waves
\begin{equation} 
\psi_{p_1,\ldots,p_n}(x_1,\ldots,x_n)=e^{-ip_1\cdot x_1} \cdots e^{-ip_n\cdot x_n} ,  
\end{equation}
where $p_a\cdot x_a=p_{a\mu}x_a^{\mu}$ and
\begin{equation}\label{e5.12} 
 p_{a0}=\pm \sqrt{{\bf p}_a^2+m_a^2} .
\end{equation}
However, not all solutions of this form are physical. From quantum field theory 
(see e.g. \cite{schweber,nikBOOK12}) one knows that only superpositions with 
positive energy (i.e. positive sign in (\ref{e5.12})) are physical, provided 
that the particles are not charged, so that their antiparticles are identical
to particles. More generally, for charged particles the positive sign in (\ref{e5.12})
describes particles while the negative sign describes antiparticles. But each
$x_a$ corresponds to either particle or antiparticle.
This means that in the wave function one may have a term
proportional to $e^{-i|E_1|x_1^0} e^{i|E_2|x_2^0}$, where the first factor 
refers to a particle and the second factor to an antiparticle.
However, it is not possible to have a superposition term of the form
$e^{-i|E_1|x_1^0}\varphi({\bf x}_1) + e^{i|E_1'|x_1^0}\varphi'({\bf x}_1)$,
because it would mean that the {\em same} particle is a superposition
of a particle and an antiparticle contribution, which quantum field theory
does not allow. 
For simplicity, in the rest of the discussion we shall only consider particles
(not antiparticles), meaning that
the energies 
\begin{equation}\label{e5.13}
 E_a=\sqrt{{\bf p}_a^2+m_a^2} 
\end{equation}
will always be positive.

 \subsection{The fundamental postulate 1: quantum proper time}
\label{SEC5.2}

Guided by analogy with the non-relativistic classical proper time (\ref{e3.34}),
non-relativistic quantum proper time (\ref{e3.35}), and 
classical relativistic proper time (\ref{e3.47.1}), we {\em postulate that the quantum relativistic
proper time $s$ is given by}
\begin{equation}\label{e5.14}
 ds^2 =\frac{1}{V_{a\nu}V_{a}^{\nu}} dX_{a\mu}dX_a^{\mu} ,
\end{equation}
where $V_{a}^{\nu}$ is given by (\ref{e5.10}).

Before studying physical consequences of this postulate, let us also write (\ref{e5.14})
in another form. Using (\ref{e5.8}), we see that (\ref{e5.5}) can be written as 
\begin{equation}\label{e5.15}
 V_{a\nu}V_{a}^{\nu}=1+\frac{2Q_a}{m_a} . 
\end{equation}
In this way, (\ref{e5.14}) can be written as
\begin{equation}\label{e5.16}
 ds^2 =\left(1+\frac{2Q_a}{m_a} \right)^{-1}  \eta_{\mu\nu}dX_a^{\mu}dX_a^{\nu} .
\end{equation}
Since $Q_a$ is a ``quantum potential'' which vanishes in the classical limit,
we see that, in this limit, the relativistic quantum proper time $ds$ given by (\ref{e5.16})
becomes the classical relativistic
proper time $d\tau$ given by (\ref{e3.47}).

In addition, we see that (\ref{e5.16}) has the form (\ref{e6}), where the anticipated 
nonlocal function $f_a(x_1,\ldots,x_n)$ is determined by the nonlocal ``quantum potential'' 
$Q_a(x_1,\ldots,x_n)$. Or viewed from another point of view, the nonlocality of (\ref{e5.14}) 
is analogous to the nonlocality of (\ref{e3.35}). 

Now comes a very important conceptual point. The classical theory of relativity asserts 
that proper time is given by $\tau$, while the present theory asserts that it is given by $s$
instead. One might object that this actually violates the theory of relativity, so in what 
sense can we claim that the present theory is relativistic? The answer is that it is relativistic, 
in the sense 
that (\ref{e5.14}) is {\em relativistic invariant}, i.e., does not depend on the choice of
spacetime coordinates. One might object that this is still not relativistic enough \cite{maudlin}, 
but we believe that Bohmian mechanics cannot be made more relativistic than that. 
Indeed, in Appendix \ref{SECA.2} we present an explicit argument for that claim,
where we argue that the requirement of spacetime equivariance excludes any local
law for the proper time. Since the classical proper time $\tau$ is local,
it implies that $\tau$ is excluded as well.

 \subsection{The fundamental postulate 2: covariant velocity law}
\label{SEC5.3}

To specify the particle trajectories we {\em postulate that particle trajectories in spacetime 
are integral curves of the field $V_{a}^{\mu}(x_1,\ldots,x_n)$.} 

To calculate these integral curves, it is useful to parameterize them with an arbitrary 
scalar parameter $\lambda$, such that the trajectories can be represented by functions 
$X_{a}^{\mu}(\lambda)$. If $\Omega(x_1,\ldots,x_n)$ is an arbitrary positive function,
one can determine the trajectories by integrating the equations
\begin{equation}\label{e5.17}
 \frac{dX_{a}^{\mu}(\lambda)}{d\lambda} =\Omega(X_1(\lambda),\ldots,X_n(\lambda)) \,
V_{a}^{\mu}(X_1(\lambda),\ldots,X_n(\lambda)) .
\end{equation}
In fact, various choices of the function $\Omega(x_1,\ldots,x_n)$ correspond 
to various choices of the parameter $\lambda$. The trajectories in spacetime do not depend
on $\lambda$ or on the choice of $\Omega$, which is most directly seen from
\begin{equation}\label{e5.18}
 \frac{dX_{a}^{\mu}} {dX_{a'}^{\mu'}} = \frac{\Omega V_{a}^{\mu}}{\Omega V_{a'}^{\mu'}}
=\frac{V_{a}^{\mu}}{V_{a'}^{\mu'}} .
\end{equation}

A particularly convenient choice is $\Omega=1$. For this choice, (\ref{e5.17}) can be written as
\begin{equation}\label{e5.19}
 \frac{dX_{a}^{\mu}}{d\lambda} = V_{a}^{\mu} .
\end{equation}
This implies 
\begin{equation}\label{e5.20}
 \frac{dX_{a\mu}}{d\lambda} \frac{dX_{a}^{\mu}}{d\lambda} = V_{a\mu} V_{a}^{\mu}
\end{equation}
i.e.
\begin{equation}\label{e5.21}
 d\lambda^2 = \frac{1}{V_{a\nu}V_{a}^{\nu}} dX_{a\mu}dX_a^{\mu} ,
\end{equation}
so comparison with (\ref{e5.14}) shows that the choice $\Omega=1$ corresponds to
$d\lambda=ds$. Thus we see that, when the particle trajectories are parameterized by the
quantum proper time (\ref{e5.14}), the equation of motion for particle trajectories is
\begin{equation}\label{e5.22}
 \frac{dX_{a}^{\mu}}{ds} = V_{a}^{\mu} ,
\end{equation}
or written in the full form
\begin{equation}\label{e5.23}
 \frac{dX_{a}^{\mu}(s)}{ds} = V_{a}^{\mu}(X_1(s),\ldots,X_n(s)) .
\end{equation}
This is a nonlocal law, because the right-hand side of (\ref{e5.23}) depends on the spacetime
positions of all particles at the same value of proper time $s$. 
Yet, this law is relativistic covariant, in the sense that it takes the same form in all
reference frames. In addition, this law does not involve any preferred foliation of spacetime. 

 \subsection{Spacetime equivariance} 
\label{SEC5.4}
 
Consider a statistical ensemble of particles moving according to the law
(\ref{e5.22}). Analogously to (\ref{e3.51}), one finds out that, in general, the ensemble
must satisfy a spacetime-equivariance equation  
\begin{equation}\label{e5.24}
\frac{\partial\rho}{\partial s} + \sum_{a=1}^{n} \partial_{a\mu}(\rho V_a^{\mu}) =0 ,
\end{equation}
where $\rho(x_1,\ldots,x_n;s)$ is the probability density that the particles have spacetime
positions $x_1,\ldots,x_n$ at the proper time $s$. 

Can we say anything more about the function $\rho(x_1,\ldots,x_n;s)$? It depends on 
how many details about the system is known. For instance, if the actual trajectories
$X_{a}^{\mu}(s)$ are known then
\begin{equation}\label{e5.25}
\rho(x_1,\ldots,x_n;s)=\delta^4(x_1-X_1(s)) \cdots \delta^4(x_n-X_n(s)) ,
\end{equation}
which leads to a generalization of (\ref{e2.10})
\begin{equation}\label{e5.26}
 \frac{\partial\rho}{\partial s}+\sum_{a=1}^{n}\partial_{a\mu}(\rho \dot{X}_a^{\mu}) =0 .
\end{equation}

In practice, however, the actual trajectories are rarely known. Hence it is more interesting to 
consider the case in which, {\it a priori}, only the wave function $\psi(x_1,\ldots,x_n)$ is known.
From the wave function, one can also determine the velocity field $V_a^{\mu}(x_1,\ldots,x_n)$
via (\ref{e5.10}). Since $V_a^{\mu}(x_1,\ldots,x_n)$ does not depend on $s$, 
with such a knowledge there is no any reason
to associate a different probability with a different value of $s$. Thus we must have
\begin{equation}\label{e5.27}
 \frac{\partial\rho}{\partial s}=0,
\end{equation}
so (\ref{e5.24}) reduces to
\begin{equation}\label{e5.28}
 \sum_{a=1}^{n}\partial_{a\mu}(\rho V_a^{\mu}) =0 .
\end{equation}
From (\ref{e5.6}) we see that this equation is satisfied if we take $\rho\propto R^2\equiv|\psi|^2$,
so up to a normalization factor we can write 
\begin{equation}\label{e5.29}
 \rho(x_1,\ldots,x_n)=|\psi(x_1,\ldots,x_n)|^2 .
\end{equation}
Indeed, this is consistent because if $\rho(x_1,\ldots,x_n;s)$ is given by (\ref{e5.29})
for some initial value of $s$, then the equivariance equation (\ref{e5.24}) provides 
that it is also so for any $s$.   

Or to summarize, the analysis above shows that the spacetime probability density (\ref{e5.29}) 
satisfies the spacetime-equivariance equation 
\begin{equation}\label{st_equivariance}
\frac{\partial|\psi|^2}{\partial s} + \sum_{a=1}^{n} \partial_{a\mu}(|\psi|^2 V_a^{\mu}) =0 .
\end{equation}

\subsection{Normalization}
\label{SECnorm}

Now let as make a few notes on normalization of $\psi$. Since $\rho$ is the probability
density, $\psi$ in (\ref{e5.29}) should be normalized such that
\begin{equation}\label{e5.30}
  \int d^4x_1 \cdots d^4x_n \, |\psi(x_1,\ldots,x_n)|^2=1 .
\end{equation}
Let us write it as 
\begin{equation}\label{e5.31}
 \int dx^0_1 \cdots dx^0_n  \, N(x^0_1,\ldots ,x^0_n) =1 ,
\end{equation}
where 
\begin{equation}\label{e5.32}
N(x^0_1,\ldots ,x^0_n)=\int d^3x_1 \cdots d^3x_n \, |\psi(x_1,\ldots,x_n)|^2 
\end{equation}
involves only spatial integrations. However, we have seen in Sec.~\ref{SEC5.1}
that $\psi$ is a positive-energy wave function. As shown in Appendix \ref{SECB.1}, 
for such a wave function (\ref{e5.32}) does not depend on $x^0_1,\ldots ,x^0_n$. 
Thus, introducing a constant $T=\int dx^0$, we see that 
\begin{equation}\label{e5.33}
 \psi(x_1,\ldots,x_n)=\frac{\tilde{\psi}(x_1,\ldots,x_n)}{\sqrt{T^n}} ,
\end{equation}
where $\tilde{\psi}$ is the wave function normalized so that
\begin{equation}\label{e5.34}
 \int d^3x_1 \cdots d^3x_n \, |\tilde{\psi}(x_1,\ldots,x_n)|^2 =1 .
\end{equation}

There is a minor technical problem related to the value of $T=\int dx^0$, 
because a natural value for it is infinite. However, it does not necessarily need to be so,
if we use a physical idea that we deal with a Universe (or a part of the Universe) that exists
only a finite time $T$. (Here ``exists'' refers to the past and future together.) 
And even for a Universe existing for an infinite time, 
we can formally work with a finite $T$ and put $T\rightarrow \infty$ at the end of calculation.
In practically all cases, nothing of actual physical interest will depend on $T$.
Indeed, we shall see in Sec.~\ref{SEC7.2} that the measurable probabilistic predictions
do not depend on $T$. 

 \subsection{Space equivariance from spacetime equivariance}
\label{SEC5.5}

A space-equivariance equation can be obtained from the spacetime-equivariance equation
(\ref{st_equivariance}), in a way similar to that in Sec.~\ref{SEC4.4.3}.
Just as assumed in Sec.~\ref{SEC4.4.3}, the quantum ensemble is stationary due to (\ref{e5.27}).
Furthermore, (\ref{e5.6}) written as
\begin{equation}\label{e5.6'}
 \partial_{a\mu}(|\psi|^2 V_a^{\mu})=0 
\end{equation}
is analogous to (\ref{e3.48_st1}).

To proceed, consider the case of an energy-eigenstate, i.e., a wave function
of the form
\begin{equation}\label{e5.35}
 \psi(x_1,\ldots,x_n)=e^{-iE_1 x^0_1}\cdots e^{-iE_n x^0_n} \, \varphi({\bf x}_1,\ldots,{\bf x}_n) . 
\end{equation}
From (\ref{e5.8}) we see that
\begin{equation}\label{e5.36}
 V^0_a=\frac{E_a}{m_a} ,
\end{equation}
which is a {\em constant}, i.e. does not depend on $x_1,\ldots,x_n$.

In practice, a realistic wave function is rarely an exact energy-eigenstate. Nevertheless,
a realistic wave function is very often an {\em approximate energy-eigenstate}, satisfying
\begin{equation}\label{e5.37}
 \Delta E_a \ll \langle E_a \rangle \equiv \bar{E}_a , 
\end{equation}
where $\Delta E_a$ is the standard deviation of energy from its average value 
$\langle E_a \rangle \equiv \bar{E}_a$. (To see how to calculate the average value and standard
deviation of energy from the wave function, see Appendix \ref{SECB.2}.)
This means that (\ref{e5.36}) can be replaced by an approximate equation 
\begin{equation}\label{e5.38}
 V^0_a \simeq \frac{\bar{E}_a}{m_a} ,
\end{equation}
implying that $V^0_a$ {\em can be approximated by a constant}.

When is the approximation (\ref{e5.37}) satisfied?
First, it is certainly satisfied in the non-relativistic limit, because in that limit
$\Delta E_a \ll m_a$ and $\bar{E}_a$ is slightly larger than $m_a$.
But the validity of (\ref{e5.37}) is actually much wider than that. In fact, we are not aware
of any actual experiment with massive particles in which one has succeeded to prepare
a pure state which would violate (\ref{e5.37}). The approximation (\ref{e5.37}) seems to be 
valid for most (if not all) practical purposes.  

Moreover, it is even possible to give a theoretical argument for the validity of (\ref{e5.37}).
In QM, there is a phenomenological charge-superselection rule \cite{pascual} saying that
a physical state cannot be in a superposition of states with different electric charges.
This superselection rule can be explained as a superselection induced by 
decoherence \cite{decoh1,giulini}. In essence, since electric field produced 
by a charge is a long-range field, a superposition of different charges will produce a
``Schr\"odinger cat'' macroscopic superposition of different electric fields, which cannot be stable
against decoherence. An analogous superselection rule is expected
for energy, for otherwise one would have macroscopic superpositions of different gravitational
fields (see also \cite{page}). 
The difference is that energy, unlike charge, has a continuous spectrum of
eigenvalues, so a sufficiently small $\Delta E_a$ cannot be completely prevented.

The wide validity of the approximation (\ref{e5.37}) motivates us to further study its consequences.
For that purpose we write (\ref{e5.6'}) as
\begin{equation}\label{e5.39}
 V_a^{0}\partial_{a0}|\psi|^2 + |\psi|^2 \partial_{a0}V_a^{0} + \partial_{ai}(|\psi|^2 V_a^{i})=0 .
\end{equation}
The first and second term in (\ref{e5.39}) are both small when (\ref{e5.37}) is valid. However, as shown 
in Appendix \ref{SECB.2}, the second term is much smaller than the first. Therefore the second term
can be neglected leading to
\begin{equation}\label{e5.40}
 V_a^{0}\partial_{a0}|\psi|^2  + \partial_{ai}(|\psi|^2 V_a^{i}) \simeq 0 ,
\end{equation}
and $V_a^{0}$ can be approximated by a constant (\ref{e5.38}).
This is analogous to (\ref{e3.48_st2}), so we can proceed in a way analogous to that in Sec.~\ref{SEC4.4.3}.
We first write (\ref{e5.40}) as
\begin{equation}\label{e5.41}
 \partial_{a0}|\psi|^2  + \partial_{ai}(|\psi|^2 v_a^{i}) \simeq 0 ,
\end{equation}
where $v_a^{i}=V_a^{i}/V_a^{0}$. Then analogously to (\ref{simple}) we find
\begin{equation}\label{simple2}
 X_a^0(s)\simeq Y_a^0+\frac{\bar{E}_a}{m_a} s,
\end{equation}
the simplicity of which implies that $x_a^0$ can be treated as a known variable.

Before proceeding with the quantitative analysis, a conceptual subtlety needs to be mentioned.
At Level I in Sec.~\ref{SEC2.1},
the microscopic proper time $s$ and the initial conditions $Y_a^0$ 
are not known to a macroscopic observer, so neither is $x_a^0$.
Therefore, for the sake of justification of the quantitative analysis that follows, 
one can think of these quantities as ``known'' 
to the microscopic particles themselves. If one finds it problematic,
we stress that it is only a temporary problem,
because in Sec.~\ref{SEC7.2} we shall provide a much better justification for such an analysis.
There a similar analysis will be applied to a macroscopic pointer, which 
a macroscopic observer can observe directly.

Having this caveat in mind, let us proceed by treating
$x_a^0$ as a known variable. Since it is known, the relevant
probability density is the conditional probability
\begin{equation}
\rho_{\rm cond}({\bf x}_1,\ldots,{\bf x}_n;x_1^0,\ldots,x_n^0)=
\frac{|\psi({\bf x}_1,x_1^0,\ldots,{\bf x}_n,x_n^0)|^2}{N} ,  
\end{equation}
where
\begin{equation}
 N=\int d^3x_1 \cdots d^3x_n \, |\psi({\bf x}_1,x_1^0,\ldots,{\bf x}_n,x_n^0)|^2 
\end{equation}
does not depend $x_1^0,\ldots,x_n^0$ because $\psi$ is a positive-energy wave function 
(see Appendix \ref{SECB.1}).
Therefore we can divide (\ref{e5.41}) by $N$ to get 
\begin{equation}\label{e5.42}
 \partial_{a0}\rho_{\rm cond}  + \partial_{ai}(\rho_{\rm cond} v_a^{i}) \simeq 0 .
\end{equation}
Finally,  
putting $x_1^0=\cdots=x_n^0=t$, summing over $a$ and using (\ref{identity}), 
we get the single-time space-equivariance equation
\begin{equation}\label{e5.43}
\frac{\partial\rho_{\rm cond}}{\partial t}   
+ \sum_{a=1}^n \partial_{ai}(\rho_{\rm cond} v_a^{i}) \simeq 0 
\end{equation}
analogous to (\ref{e3.48_st6}).

The space-equivariance equation (\ref{e5.43}) is not covariant. So in which Lorentz frame
is it satisfied? For most wave functions, it is not satisfied in any Lorentz frame exactly.
Yet, it is satisfied approximately in all Lorentz frames in which (\ref{e5.37}) is a good approximation.

\section{Further generalizations}
\label{SEC6}

So far we have considered free massive particles without spin, and the number of particles was
fixed. This is certainly not the most general case, so in this section we discuss
how to make the appropriate generalizations. Most of the results, however, of the present section are not new,
so the presentation in this section is very brief, written only for the sake of completeness,
with references to the work with more details. The main new result of this section
is the argument in Sec.~\ref{SEC6.2} that massless particles do not have trajectories. 

 \subsection{Spin and other discrete degrees of freedom}
\label{SEC6.1}

For particles with spin, the $n$-particle wave function has the form $\psi_{l_1,\ldots,l_n}(x_1,\ldots,x_n)$, 
where $l_a$ are discrete labels. Additional discrete labels may be present when the theory has some 
additional internal symmetry, such as SU(2) or SU(3) in the Standard Model of elementary particles.
To simplify notation, we introduce one collective label $A=(l_1,l_2,\ldots )$ and denote the wave function as
$\psi_A(x_1,\ldots,x_n)$. Then all components of the wave function can be represented by a column
\begin{equation}
 \psi=\left( 
\begin{array}{c} 
 \psi_1 \\ \psi_2 \\ \vdots
\end{array}
\right) . 
\end{equation}

With this compact notation, it is straightforward to 
make the appropriate generalizations of all results in Sec.~\ref{SEC5}.
In particular, all expressions of the form $\psi^* O \psi$ should be replaced by $\psi^{\dagger} O \psi$
\cite{nikBOOK12}.
For instance, (\ref{e5.29}) written as $\rho=\psi^*\psi$ generalizes to
\begin{equation}\label{dag1}
 \rho=\psi^{\dagger}\psi ,
\end{equation}
while the conserved current (\ref{e5.2}) generalizes to a new conserved current
\begin{equation}\label{e5.2dag}
j_{a\mu} = \frac{i}{2m_a} \, \psi^{\dagger} \stackrel{\leftrightarrow\;\;}{\partial_{a\mu}} \psi .
\end{equation}
The velocity field (\ref{e5.10}) generalizes to   
\begin{equation}\label{e5.10dag}
 V_{a}^{\mu}=\frac{j_a^{\mu}}{\psi^{\dagger}\psi} ,
\end{equation}
which determines particle trajectories as in (\ref{e5.22}).

Another illuminating notation is to write (\ref{dag1}) and (\ref{e5.2dag}) as
 \begin{equation}\label{tr1}
 \rho={\rm Tr}_{\rm discr}\psi^*\psi ,
\end{equation}
\begin{equation}\label{e5.2tr}
j_{a\mu} = \frac{i}{2m_a} \, {\rm Tr}_{\rm discr} \psi^{*} \stackrel{\leftrightarrow\;\;}{\partial_{a\mu}} \psi ,
\end{equation}
where ${\rm Tr}_{\rm discr}$ denotes trace-out of the all discrete degrees of freedom.

When (\ref{dag1}) and (\ref{e5.2dag}) are applied to spin-$\frac{1}{2}$, an additional clarification
is needed concerning the transformation properties of (\ref{dag1}) and (\ref{e5.2dag}).
For simplicity, consider the single-particle case, so that (\ref{e5.2dag}) reduces to
\begin{equation}\label{curnoldn}
j_{\mu} = \frac{i}{2m} \, \psi^{\dagger} \stackrel{\leftrightarrow\;}{\partial_{\mu}} \psi .
\end{equation}
From known transformation properties of spinors under Lorentz transformations
\cite{bd1}, one might naively conclude that (\ref{curnoldn}) does not transform as a
vector, so that one should take the Dirac current $j'_{\mu}=\bar{\psi}\gamma_{\mu}\psi$ instead.
However, this is not really true \cite{nikIJMPA10,nikBOOK12,nik_dirac}. 
The standard spinor-transformation properties \cite{bd1} cannot be generalized to curved
spacetime, so for general purposes it is more convenient to redefine the transformation properties
of spinors and Dirac matrices such that $\psi$ transforms as a scalar and $\gamma_{\mu}$ 
as a vector under coordinate transformations \cite{birdel,parker}. Such a redefinition
of transformations does not alter the vector transformation properties of 
the Dirac current, but implies that (\ref{curnoldn}) also transforms as a vector.
Likewise, since $\psi$ transforms as a scalar, it follows that (\ref{dag1}) transforms as a scalar too.
A particularly simple and pedagogic exposition of these ideas is presented in \cite{nik_dirac}.

 \subsection{Massless particles}
\label{SEC6.2}

So far we have considered only massive particles. Can we generalize our results to the 
massless particles as well? An immediate problem is that equations such as (\ref{e5.2}) and (\ref{e5.8})
are not well defined for $m_a=0$. One possibility is to reformulate the theory such that these equations
do not have the factor of mass, but then $s$ in (\ref{e5.14}) will no longer have the dimension of time,
so our interpretation of $s$ as the fundamental microscopic proper time would not make sense.
Another possibility is to take some {\em ad hoc} mass scale $M$ in equations 
such as (\ref{e5.2}) and (\ref{e5.8}).
However, one is not allowed to modify the wave equation (\ref{e5.1}), which
for $m_a^2=0$ leads to
\begin{equation}\label{HJm=0}
 (\partial_{a\mu}S)(\partial_{a}^{\mu}S)-\frac{\partial_{a\mu}\partial_{a}^{\mu}R}{R}=0  
\end{equation}
instead of (\ref{e5.5}). The second term in (\ref{HJm=0}) vanishes in the classical limit,
so (\ref{e5.8}) with $m_a\rightarrow M$ leads to $V_{a\mu}V_{a}^{\mu}=0$ in the classical limit,
which is not compatible with the requirement that (\ref{e5.14}) in the classical limit 
should become $ds^2=dX_{a\mu}dX_a^{\mu}$.

The simplest resolution of this problem is to propose 
that massless particles do not have trajectories at all, i.e. that massless 
particles are not ontological. 
Indeed, classical massless particles experience no proper time, which is compatible with the idea that they  
might not really exist. But if classical massless particles do not exist, it seems perfectly reasonable
to assume that quantum (Bohmian) particles do not exist as well. We do not claim that this is necessary
the only logical possibility, but it seems to be the simplest and most natural possibility
compatible with our interpretation of $s$ as the fundamental proper time that, in the classical limit,
should reduce to the standard classical proper time.

But if massless particles do not exist, then how can they be observed? As we shall see in Sec.~\ref{SEC7.2},
the behavior of the massless-particle detector can be explained 
even if the massless particle itself does not exist. For that purpose, 
it is sufficient that the detector (made up
of massive particles) exists, provided that the wave function for all particles, including the 
massless ones, is given as well.

In this way, the treatment of massless particles is analogous to the treatment of spin. Analogously 
to (\ref{tr1}) and (\ref{e5.2tr}), in the case of massless particles we have
 \begin{equation}\label{tr1m}
 \rho={\rm Tr}_{m=0}\psi^*\psi ,
\end{equation}
\begin{equation}\label{e5.2trm}
j_{a\mu} = \frac{i}{2m_a} \, {\rm Tr}_{m=0} \psi^{*} \stackrel{\leftrightarrow\;\;}{\partial_{a\mu}} \psi ,
\end{equation}
where ${\rm Tr}_{m=0}$ denotes trace-out of all degrees of freedom associated with massless particles
and the label $a$ refers to a massive particle.
For instance, if we have a two-particle wave function $\psi(x_1,x_2)$ where the coordinate $x_1$ refers to the 
massless particle, then (\ref{tr1m}) and (\ref{e5.2trm}) in the explicit form read
\begin{equation}\label{tr1me}
 \rho(x_2)=\int d^4x_1 \, \psi^*(x_1,x_2)\psi(x_1,x_2) ,
\end{equation}
\begin{equation}\label{e5.2trme}
j_{2\mu}(x_2) = \frac{i}{2m_2} \int d^4x_1 \,  
\psi^{*}(x_1,x_2) \stackrel{\leftrightarrow\;\;}{\partial_{2\mu}} \psi(x_1,x_2) .
\end{equation}

 \subsection{Particles in classical background fields}
\label{SEC6.3}
 
So far we have only considered the free Klein-Gordon equation. Here we briefly outline
how this generalizes when the particle moves in a classical background electromagnetic or gravitational
field. Essentially, all one has to do is to replace the derivative $\partial_{\mu}$ with the appropriate
covariant derivative. For simplicity we do the analysis for spin-0 particles, but a similar 
analysis can be done for any spin. 

It should also be noted that a classical background field is only an approximation,
because a full theory requires quantum fields (which are very briefly outlined in
Sec.~\ref{SEC6.4}). For example, for strong background fields it may happen that a positive-energy
wave function evolves into a wave function which is no longer a positive-energy one.
When this happens, it means that particle creation takes place. Particle creation can be correctly 
described only by quantum field theory, implying that in this case the approximation 
is no longer applicable.    

     \subsubsection{Background electromagnetic field}
\label{SEC6.3.1}

The effects of an electromagnetic background $A^{\mu}(x)$ on the 
wave function of the particle with charge
$e$ are described by the gauge-covariant derivative
\begin{equation}\label{KGem}
 D_{\mu}=\partial_{\mu}+ieA_{\mu}(x) .
\end{equation}
The appropriate generalization of (\ref{antisimder}) is 
\begin{equation}\label{antisimder2}
A \!\stackrel{\leftrightarrow\;}{D_{\mu}}\! B \equiv
A D_{\mu}B - (D^*_{\mu}A) B ,
\end{equation}
where $D^*_{\mu}=\partial_{\mu}-ieA_{\mu}$.

In this way, the Klein-Gordon equations (\ref{e5.1}) generalize to
\begin{equation}\label{KGA}
(D_a^{\mu} D_{a\mu} +m_a^2)\psi(x_1,\ldots,x_n)=0,
\end{equation}
where
\begin{equation}
 D_{a\mu}=\partial_{a\mu}+ie_aA_{\mu}(x_a) .
\end{equation}
The currents (\ref{e5.2}) generalize to 
\begin{equation}\label{e5.2em}
j_{a\mu} = \frac{i}{2m_a} \, \psi^* \stackrel{\leftrightarrow\;\;}{D_{a\mu}} \psi ,
\end{equation}
and (\ref{KGA}) implies that these currents are conserved
\begin{equation}
 \partial_{a\mu} j_a^{\mu} = 0. 
\end{equation}
The velocity field is
\begin{equation}
 V_a^{\mu}=\frac{j_a^{\mu}}{\psi^*\psi} ,
\end{equation}
and the corresponding particle velocities given by
\begin{equation}
 \frac{dX_a^{\mu}}{ds}=V_a^{\mu}
\end{equation}
can also be written as
\begin{equation}\label{param1.cl}
 \frac{dX^{\mu}_a}{ds} = -\frac{P_a^{\mu}}{m_a} ,
\end{equation}
where
\begin{equation}
 P_{a\mu}(x_1,\ldots,x_n)\equiv \partial_{a\mu}S(x_1,\ldots,x_n)+e_aA_{\mu}(x_a) ,
\end{equation}
and $S$ is defined by $\psi=Re^{iS}$.

     \subsubsection{Background gravitational field}
\label{SEC6.3.2}

In a classical background gravitational field, 
the Minkowski metric $\eta_{\mu\nu}$ is replaced by the curved metric
$g_{\mu\nu}(x)$. The Klein-Gordon equations (\ref{e5.1}) generalize to
\begin{equation}\label{KGg}
(\nabla_a^{\mu} \nabla_{a\mu} +m_a^2)\psi(x_1,\ldots,x_n)=0,
\end{equation}
where $\nabla_{a\mu}$ is the general-covariant derivative \cite{carroll}.
The currents (\ref{e5.2}) become 
\begin{equation}\label{e5.2g}
j_{a\mu} = \frac{i}{2m_a} \, \psi^* \stackrel{\leftrightarrow\;\;\;\;}{\nabla_{a\mu}} \psi ,
\end{equation}
and (\ref{KGg}) implies that these currents are covariantly conserved
\begin{equation}
 \nabla_{a\mu} j_a^{\mu} = 0. 
\end{equation}
The velocity field is
\begin{equation}
 V_a^{\mu}=\frac{j_a^{\mu}}{\psi^*\psi} ,
\end{equation}
and the corresponding particle velocities are
\begin{equation}
 \frac{dX_a^{\mu}}{ds}=V_a^{\mu} .
\end{equation}
The quantum proper time (\ref{e5.14}) generalizes to
\begin{equation}\label{e5.14g}
 ds^2 =\frac{1}{V_{a\nu}V_{a}^{\nu}} \, g_{\mu\nu}dX_a^{\mu} dX_a^{\nu},
\end{equation}
where $d\tau_a^2=g_{\mu\nu}dX_a^{\mu} dX_a^{\nu}$ is the classical proper time 
in the gravitational background.

 \subsection{Quantum field theory}
\label{SEC6.4}

So far we have only considered the case in which the number $n$ of particles is fixed. But in quantum field theory
(QFT) $n$ may not be fixed. Instead, a general QFT state has a form
\begin{equation}
 |\Psi\rangle=\sum_{n=0}^{\infty} c_n |n\rangle , 
\end{equation}
where $|n\rangle$ is an (appropriately normalized) state with $n$ particles.

With each term $c_n |n\rangle$ one can associate an unnormalized $n$-particle wave function
$\tilde{\Psi}_n(x_{n1},\ldots,x_{nn})$ \cite{nikIJMPA10,nikBOOK12}. Then the simplest way to introduce
the total wave function is \cite{schweber,nik10}
\begin{equation}
  \Psi(x_{11},x_{21},x_{22},\ldots)=\left[ 
\begin{array}{l} 
 \tilde{\Psi}_0 \\ \tilde{\Psi}_1(x_{11}) \\ \tilde{\Psi}_2(x_{21},x_{22}) \\ \vdots
\end{array}
\right] .
\end{equation}
The associated probability density and current are
\begin{equation}
  \rho=\Psi^{\dagger}\Psi ,
\end{equation}
\begin{equation}
 j_{na\mu} = \frac{i}{2m_{na}}   
\Psi^{\dagger} \stackrel{\leftrightarrow\;\;\;\;}{\partial_{na\mu}} \Psi ,
\end{equation}
for $n=1,\ldots,\infty$ and $a=1,\ldots,n$. From those density and current, Bohmian particle
trajectories are calculated in a way similar to that of a fixed number of particles \cite{nikIJMPA10,nikBOOK12}.
Even though the particle trajectories are continuous and there may be an infinite number of them, 
a particle detector behaves as if a finite number of particles is created or destructed, 
in agreement with observations \cite{nikIJMPA10,nikBOOK12,nik10}.

\section{Quantum measurement}
\label{SEC7}

 \subsection{Non-relativistic quantum measurement and a dark-matter analogy}
\label{SEC7.1}

The quantum theory of measurement in non-relativistic Bohmian mechanics
is a well-developed subject \cite{bohm,book-bohm,book-hol,book-durr,apl-book,durr92}. 
Yet, there is one aspect of it
which, in our opinion, has not been explained sufficiently clearly. 

Let us briefly explain it. Let 
\begin{equation}\label{e7.1}
\psi({\bf x},t)=\sum_b c_b \psi_b({\bf x},t) 
\end{equation}
be the initial wave function of the system that will be measured, and
let after the measurement 
\begin{equation}\label{e7.2}
 \Psi({\bf x},{\bf y},t)=\sum_b c_b \psi_b({\bf x},t) \phi_b({\bf y},t)
\end{equation}
be the wave function describing the entanglement with the measuring apparatus. 
The wave functions $\phi_b({\bf y},t)$ are the 
well-localized pointer states of the apparatus. They do not mutually overlap
in the ${\bf y}$-space, in the sense that
\begin{equation}
\phi_b({\bf y},t) \phi_{b'}({\bf y},t) \simeq 0 \;\;\; {\rm for} \;\;\; b\neq b' . 
\end{equation}
Even though Bohmian particles have well defined positions $X^i$,
it is asserted that they cannot be measured with a precision better than 
the standard deviation $\Delta x^i$ of the wave function. 
But {\em which} wave function, $\Psi$ or $\psi_b$? It is asserted
that this refers to the wave function $\psi_b$. But why? Because 
the actual position ${\bf Y}$ of the macroscopic pointer  
is {\em perceived directly} 
(i.e. at Level I of Sec.~2.1), 
so one knows which of the non-overlapping channels is actually filled. 
If the filled channel
is $\phi_b$, then one can use the Bohmian theory of motion (at Level II of Sec.~2.1)
to conclude that the measured particle with position ${\bf X}$
will behave as if the initial wave function (\ref{e7.1}) collapsed 
to $\psi_b$ in (\ref{e7.2}). For that reason, $\psi_b$ is referred to as {\em effective wave function}.

So far so good. But now we want to address some questions which the explanation above 
still leaves unanswered. Why exactly the actual position
${\bf Y}$ of the macroscopic apparatus is perceived directly? 
And what exactly does it even 
{\em mean} to be ``perceived directly''? In addition, since that determines
which of the channels $\phi_b$ is actually filled, this means that 
$Y^i$ is determined with a precision better than $\Delta y^i$ associated
with the {\em true} wave function (\ref{e7.2}). Hence, if we could ``directly perceive''
${\bf X}$ just as we can ${\bf Y}$, then we could have a direct
empirical proof that Bohmian particle positions really exist, just as we do
have a direct empirical proof that macroscopic pointers really exist.
Thus, it is important to understand what exactly makes ${\bf X}$ 
different from ${\bf Y}$. Is it just that ${\bf Y}$ describes something
``big'', or is there a more profound explanation? In our opinion, 
those questions have not been answered sufficiently clearly in the literature.       

To find a more appealing answer to those questions, we find illuminating
to explore an analogy with a well-known paradigm from a totally different 
branch of physics -- astrophysics.
It is known that about 80\% of all matter in the Universe is the so-called
{\em dark matter} \cite{darkmatter}, a kind of matter that does not produce light and therefore
cannot be seen. But if it cannot be seen, then how do we know that it exists?
In fact we don't! Or at least, not with the same level of confidence 
with which we know that the luminous matter (such as stars) exists. Indeed,
there are some alternative theories \cite{darkmatter} which suggest that dark matter might 
not exist at all. But then again, if it might not actually exist,
then why do most astrophysicists still think
that it probably exists? The point is that {\em dark matter
is not perceived directly, but is perceived indirectly}. 

But what exactly does it mean that dark matter is perceived {\em indirectly}?
By observing (Level I) the motion of luminous matter in galaxies 
and using the known theory of gravity (at Level II.A), one finds out 
that the observed motion of luminous matter cannot be explained by the luminous matter
alone. If the used theory of gravity is correct, then, 
to explain the observed motion of luminous matter, there must be
some additional matter in galaxies. And since it isn't seen, it must be dark.

Thus the observation of dark matter is indirect, in the sense that we really observe
luminous matter and from it make inference about dark matter. However, strictly speaking,
the observation of luminous matter is also indirect, because we do not really see 
luminous matter itself. Instead, what we really see is the light that comes to our telescopes,
and from this light we make inference about luminous matter. So what exactly 
makes the observation of luminous matter more direct than the observation of
the dark one?   

The point is that the interaction between light (which we really see) and luminous matter
(about which we infer from the light seen) is {\em very localized}. Namely, luminous
matter influences the behavior of light only at the place of the luminous matter itself.
(A nit-picking reader might object that there is also a not-so-localized interaction
based on gravitational lensing, but for most purposes this effect may be neglected.)
Thus, from observed light, one can easily and precisely determine 
the {\em position} of the luminous matter.  
On the other hand, the influence of dark matter on luminous matter is based on the
{\em long-range} gravitational interaction. In this way luminous matter may be influenced
by dark matter positioned very far away from the luminous one, so, by determining the position of
luminous matter, one cannot so easily and precisely determine the position of dark matter.
And since humans tend not to be absolutely convinced that something exists until
they are able to detect the exact place where it exists, the detection of dark matter
based on detection of luminous matter is considered less convincing than the detection
of luminous matter itself.  

Now, equipped with this astro-epistemological insight, we can turn back to Bohmian mechanics.
At the fundamental microscopic level, the interaction between the Bohmian particles
is highly nonlocal. This is why influence of particle A on particle B cannot be used
to easily and precisely determine the position of particle A, even if the position of 
the ``detector'' particle B is somehow known. And this is why there is no
direct experimental evidence that microscopic Bohmian particles really exist.
(A modified Bohmian theory which would include 
a local contact interaction between the microscopic particles
would lead to measurable predictions that could easily be tested.)

On the other hand, macroscopic objects in Bohmian mechanics have a different behavior.
Owing to the decoherence \cite{decoh1,decoh2}, the wave packets are localized in space,
while owing to the Ehrenfest theorem \cite{ballentine}, 
the wave packets (within which the motion of particles is confined) move according to the classical laws. 
Unlike the fundamental microscopic laws, the macroscopic classical laws are local. 
Thus, it is this macroscopic locality which is responsible for the fact that we can
easily and precisely determine the position of the apparatus pointer ${\bf Y}$, and not
the position ${\bf X}$ of the microscopic particle. This is why ${\bf Y}$, and not ${\bf X}$,
is ``directly perceived''.    

In other words, microscopic Bohmian particles are analogous to dark matter, 
macroscopic objects made of particles are analogous to luminous matter, 
and quantum nonlocality is analogous to the long-range gravitational interaction.

 \subsection{General theory of quantum measurement in relativistic Bohmian mechanics}
\label{SEC7.2}

\subsubsection{Spacetime probability for the macroscopic pointer}

Let the wave function $\psi(x)$ before measurement be expanded as
\begin{equation}\label{e7.3}
\psi(x)=\sum_b c_b \psi_b(x) , 
\end{equation}
where 
\begin{equation}
 \sum_b |c_b|^2=1
\end{equation}
and $\psi_b(x)$ are some wave functions normalized so that
\begin{equation}\label{e7norm}
\int d^4x \, |\psi_b(x)|^2 =1.
\end{equation}
Let the initial wave function of the measuring apparatus be $\phi_0(y)$, and let
the interaction with the measuring apparatus be such that
the evolution from times before to times after measurement has the form
\begin{equation}
 \psi_b(x)\phi_0(y) \rightarrow \psi_b(x)\phi_b(y) ,
\end{equation}
where the pointer states $\phi_b(y)$ are well-localized in the ${\bf y}$-space, i.e. satisfy
\begin{equation}\label{nooverlap}
\phi_b({\bf y},y^0) \phi_{b'}({\bf y},y^0) \simeq 0 \;\;\; {\rm for} \;\;\; b\neq b' . 
\end{equation}
Then for $x^0$ and $y^0$ after the measurement, 
the linearity of evolution implies that the total wave function is
\begin{equation}\label{e7.4}
 \Psi(x,y)=\sum_b c_b \psi_b(x) \phi_b(y) .
\end{equation}
This is a decohered wave function, in the sense that (\ref{nooverlap}) implies
that the joint probability density 
\begin{equation}\label{e7.5}
\rho(x,y)= |\Psi(x,y)|^2 \simeq \sum_b |c_b|^2 |\psi_b(x)|^2 |\phi_b(y)|^2 
\end{equation}
does not contain the interference terms between different terms in (\ref{e7.4}).   

The pointer position $Y$ is a macroscopic quantity which interacts with the observer
in a local way. The interaction is local in both space and time, so
from the analysis in Sec.~\ref{SEC7.1} it follows that $Y$ can be
``perceived directly''. The position $X$, on the other hand, cannot be directly observed,
so one is interested in the marginal probability density of the only directly observable quantity 
\begin{equation}\label{e7.6}
 \rho(y)=\int d^4x \, \rho(x,y) \simeq \sum_b |c_b|^2 |\phi_b(y)|^2 .
\end{equation}

Alternatively, if $x$ is a position associated with a wave function of a massless particle
(which does not have a trajectory according to Sec.~\ref{SEC6.2}), then
(\ref{e7.6}) is obtained in a different way, not as a marginal probability, but from
the fundamental formula (\ref{tr1me}). 

\subsubsection{Space probability for the macroscopic pointer}

Let us recapitulate what we have achieved so far. First, we have obtained
the probability (\ref{e7.6}) for the {\em macroscopic pointer}, an object which we really
see in the laboratory. That is good. Second, the result (\ref{e7.6}) is covariant.
This is more than good; this is a real progress compared to non-relativistic
Bohmian mechanics. Third, (\ref{e7.6}) describes the spacetime probability, not 
the space probability. This is not so good, because this is not how 
we are used to think about the macroscopic world.

What we want is to obtain a space probability density from the spacetime probability
density (\ref{e7.6}). But this is not really difficult; we can 
do that in three little steps. 
In the first step, we 
introduce a {\em macroscopic clock}. The macroscopic clock cannot measure 
the fundamental microscopic proper time $s$ (see also Appendix \ref{SECC}). 
Instead, it measures
the macroscopic classical proper time $\tau$ given by
\begin{equation}
d\tau^2 =\eta_{\mu\nu}dY^{\mu}dY^{\nu} .
\end{equation}   
In the second step, we recall from Sec.~\ref{SEC2.1'} that $\tau$ is naturally 
interpreted as a known variable, not subject to a probabilistic 
description. 

In the third step we use the fact that (\ref{e7.6}) is covariant,
implying that we can choose any coordinates we want. So we choose
some coordinates in which the 3-velocity of the pointer
is small compared to the velocity of light, implying that
\begin{equation}
dY^0 \simeq d\tau .
\end{equation}
Since $\tau$ is naturally interpreted as a known variable, 
this means that, in these coordinates, $y^0$ is also naturally interpreted 
as an approximately known variable, not subject to a probabilistic description.
 
Therefore, for practical purposes, the relevant probability density
is not (\ref{e7.6}), but the conditional probability 
\begin{equation}\label{e7.7}
\rho_{\rm cond}({\bf y};y^0) = \frac{\rho({\bf y},y^0)}{N} ,
\end{equation}
where $N=\int d^3y \, \rho({\bf y},y^0)$ does not depend on $y^0$ due to the positivity of energy
(see Appendix \ref{SECB.1}). 
In this way (\ref{e7.6}) leads to
\begin{equation}\label{e7.8}
 \rho_{\rm cond}({\bf y};y^0) \simeq \sum_b |c_b|^2 |\tilde{\phi}_b({\bf y},y^0)|^2 ,
\end{equation}
where
\begin{equation}\label{e7.9}
 \tilde{\phi}_b({\bf y},y^0)=\frac{\phi_b({\bf y},y^0)}{\sqrt{N}} 
\end{equation}
is normalized so that
\begin{equation}\label{e7.10}
 \int d^3y \, |\tilde{\phi}_b({\bf y},y^0)|^2 =1 .
\end{equation}

Finally, let ${\rm supp}\;\tilde{\phi}_b$ be the support of $\tilde{\phi}_b({\bf y},y^0)$, i.e.
the region in ${\bf y}$-space within which $\tilde{\phi}_b({\bf y},y^0)$ is not negligible.
Eq.~(\ref{e7.8}) implies that the probability to find ${\bf y}$ in the support
of $\tilde{\phi}_b$ at time $y^0$ is
\begin{equation}\label{e7.11}
p_b=\int_{{\rm supp}\;\tilde{\phi}_b} d^3y \, \rho_{\rm cond}({\bf y};y^0) \simeq |c_b|^2 .
\end{equation}
This coincides with the probabilistic predictions of standard QM and does not depend
on the value of $T$ in Sec.~\ref{SECnorm}.

\subsubsection{Space equivariance for the macroscopic pointer}

Now let us show that the space probability density (\ref{e7.8}) satisfies a space-equivariance equation.
When $Y$ is in the $\phi_b(y)$-branch, then the decohered wave function (\ref{e7.4}) implies
that $\phi_b(y)$ is the effective wave function for $Y$. The approximation (\ref{e5.37}) 
is certainly very well satisfied for $\phi_b(y)$,
so we can apply the results of Sec.~\ref{SEC5.5} to conclude that
the space probability density 
$|\tilde{\phi}_b({\bf y},y^0)|^2$ satisfies a space-equivariance equation
\begin{equation}\label{e7.12}
\frac{\partial|\tilde{\phi}_b|^2}{\partial y^0} + \partial_{i}(|\tilde{\phi}_b|^2 v^{i}) \simeq 0 
\end{equation}
as a very good approximation. By multiplying this equation with $|c_b|^2$, summing over $b$ and
using (\ref{e7.8}), we finally obtain 
\begin{equation}\label{e7.spemacro}
\frac{\partial \rho_{\rm cond}}{\partial y^0} + \partial_{i}(\rho_{\rm cond} v^{i}) \simeq 0 ,
\end{equation}
which is the space-equivariance equation for (\ref{e7.8}).

Here we have used results of Sec.~\ref{SEC5.5}, which were performed in some Lorentz frame
in which the approximation (\ref{e5.37}) was valid. Since the Lorentz frame 
is characterized by a constant velocity of the frame,
(\ref{e7.spemacro}) is certainly correct for the case when the variation of the pointer velocity 
is small compared to the velocity of light. But what if the localized branch $\phi_b(y)$
changes velocity by an amount
comparable to the velocity of light? Or what if the velocity of one branch 
differs from that of another branch by an amount
comparable to the velocity of light?

Even in such a more general situation, each branch $\phi_b(y)$ is well-localized 
around some macroscopic timelike trajectory in spacetime. Since the theory is covariant
not only under Lorentz transformation, but under general coordinate transformations  as well
(see also Sec.~\ref{SEC6.3.2}), one can choose some curved coordinates which look like local
Lorentz coordinates along each of these timelike trajectories (see e.g. \cite{mtw,nikPRA00}).
When expressed in such coordinates, (\ref{e7.spemacro}) is correct even for such a more general
situation.

 \subsection{Examples}
\label{SEC7.3}

     \subsubsection{Measurement of position in space}
\label{SEC7.3.1}

In relativistic QM, the space-position operator can be defined, but not in a relativistic covariant way
\cite{newton,philips}. This means that a wave function (supposed to be an eigenstate of the position operator)
localized in space in one Lorentz frame cannot be localized in space in all Lorentz frames.

This, however, does not mean that Bohmian mechanics cannot be formulated in a relativistic covariant form.
As we have seen in this paper, it can.
All this means is that, when the particle space-position is measured, it is always measured in some particular 
Lorentz frame. Or more precisely, it means that the wave function $\psi_b(x)$ in (\ref{e7.4})
is localized in ${\bf x}$-space in one particular Lorentz frame, at some particular value 
of time $x^0$. When $\psi_b(x)$ is such an appropriately localized wave function, then 
Sec.~\ref{SEC7.2} describes the measurement of position in space. 

Note also that in relativistic Bohmian mechanics, the position ${\bf x}$ is not a preferred observable.
(If it was, then the theory would not be relativistic covariant.) The probability density of 
particle position in space is not always given by $|\psi|^2$, which is related to the fact that
the space-equivariance equation is not always satisfied, except under certain approximations
studied in Sec.~\ref{SEC5.5}. Instead, the preferred quantity (though not strictly an observable 
in a technical sense, see Sec.~\ref{SEC7.3.2}) is the {\em spacetime} position $x$, the probability 
$|\psi|^2$ of
which satisfies the spacetime-equivariance equation {\em exactly}.

Nevertheless, when the space position is {\em measured}, then the final result (\ref{e7.11}), applied to the
case in hand, reveals that the probability to get a given value of ${\bf x}$ 
as the result of measurement (i.e., as a value pointed by the macroscopic pointer) 
is given by $|\psi|^2$. And this is true even if that refers to a massless particle, which, according
to Sec.~\ref{SEC6.2}, does not have a position at all.

     \subsubsection{Measurement of time}
\label{SEC7.3.2}

Technically, time is not an observable in QM \cite{pauli,nikFP07}. The simplest way to understand this 
\cite{nikIJQI09} is to note that a superposition of plane waves with positive energies cannot give
$\delta(t)$, or any other wave function localized in time (see also Appendix \ref{SECB.1}).
Thus there are no physical eigenstates $|t\rangle$ of the time operator $\hat{t}$, and 
without eigenstates, one cannot construct the operator $\hat{t} =\sum_t t |t\rangle \langle t|$ either.
(For the same reason, the spacetime position is also not an operator, because there are no
physical wave functions localized in spacetime.)

Nevertheless, it does not mean that time cannot be measured. Time is measured by a macroscopic clock,
which is why it is best viewed as a classical quantity. As we have already discussed in Sec.~\ref{SEC7.2},
a macroscopic clock can measure $Y^0$ of the macroscopic pointer, which is nothing but a 
measurement of time. 

     \subsubsection{Measurement of velocity}
\label{SEC7.3.3}

In general, at some points in spacetime, 
the velocity (\ref{e5.22}) may be superluminal,
i.e. higher than the velocity of light. This, however, is not in contradiction with relativity.
As discussed in more detail in \cite{nikBOOK12}, at these points the effective mass squared 
is negative, which corresponds to a tachyon -- a fully relativistic object moving faster than light.

However, a particle moving with the velocity of light has never been observed. So is such a 
superluminal behavior in contradiction with observations? No it isn't! When the velocity is measured,
then the {\em measured} velocity cannot be superluminal.

This can be seen at two levels. First, at the macroscopic level, the observed velocity
of a macroscopic object is determined by the velocity of the localized wave packet, 
the motion of which is essentially classical so that it never exceeds the velocity of light.

Second, at the microscopic level, the measurement of velocity corresponds to the case
in which $\psi_b(x)$ in (\ref{e7.4}) are eigenstates of the velocity operator. 
The velocity eigenstates are the same as the momentum eigenstates 
\begin{equation}\label{ep}
\psi_p(x)=e^{-ip\cdot x} ,
\end{equation}
where, due to (\ref{e5.13}), the 4-momentum $p$ satisfies $p\cdot p=m^2$, which cannot be negative.
Consequently, the effective wave function described by (\ref{ep}) leads to a Bohmian velocity
which cannot exceed the velocity of light. 
The Bohmian velocity may exceed the velocity of light when the effective wave function
is a superposition of different plane waves of the form of (\ref{ep}), but such a situation
does not correspond to a situation in which the microscopic velocity is accurately measured.    

     \subsubsection{Measurement of nonlocal correlations}
\label{SEC7.3.4}

One of the main motivations to introduce a relativistic covariant version of 
Bohmian mechanics in the first place,
is to explain how can the measurable nonlocal quantum correlations be compatible with relativity.
So let us present a detailed explanation here.

The equations in Sec.~\ref{SEC7.2} are written down for the case of measurement of one particle.
On the other hand, to measure nonlocal correlations, at least two particles need to be measured.
Nevertheless, it is not difficult to rewrite all the equations of Sec.~\ref{SEC7.2}
to describe the measurement of an arbitrary number of particles.

For simplicity, let us do that for two particles. Instead of an expansion
in terms of the single-particle
wave functions $\psi_b(x)$, we have an expansion in terms of the two-particle wave functions
\begin{equation}
 \psi_{b_1b_2}(x_1,x_2)=\zeta_{b_1}(x_1) \xi_{b_2}(x_2) .
\end{equation}
Similarly, instead of the single measuring apparatus with pointer states $\phi_b(x)$,
we have two measuring apparatuses with pointer states of the form
\begin{equation}
 \Phi_{b_1b_2}(y_1,y_2)=\phi_{b_1}(y_1) \varphi_{b_2}(y_2) .
\end{equation}
Hence, instead of (\ref{e7.3}), the initial entangled state before measurement is
\begin{equation}\label{e7.3-2}
\Psi(x_1,x_2)=\sum_{b_1,b_2} c_{b_1b_2} \, \zeta_{b_1}(x_1) \xi_{b_2}(x_2) . 
\end{equation}
Likewise, after the measurement, instead of (\ref{e7.4}) we have 
\begin{equation}\label{e7.4-2}
 \Psi(x_1,x_2,y_1,y_2)= \sum_{b_1,b_2} c_{b_1b_2} \, \zeta_{b_1}(x_1)\phi_{b_1}(y_1) \otimes 
\xi_{b_2}(x_2) \varphi_{b_2}(y_2) ,
\end{equation}
so instead of (\ref{e7.5}) we obtain
\begin{eqnarray}\label{e7.5-2}
& \rho(x_1,x_2,y_1,y_2)= |\Psi(x_1,x_2,y_1,y_2)|^2 &
\nonumber \\ 
& \simeq \displaystyle\sum_{b_1,b_2} |c_{b_1b_2}|^2 \, |\zeta_{b_1}(x_1)|^2 |\phi_{b_1}(y_1)|^2  
|\xi_{b_2}(x_2)|^2 |\varphi_{b_2}(y_2)|^2 , &
\end{eqnarray}
and instead of (\ref{e7.6})  
\begin{equation}\label{e7.6-2}
 \rho(y_1,y_2)\simeq \sum_{b_1,b_2} |c_{b_1b_2}|^2 \, |\phi_{b_1}(y_1)|^2|\varphi_{b_2}(y_2)|^2  .
\end{equation}
Thus, in a completely analogous way, instead of (\ref{e7.11}) we finally obtain
\begin{equation}\label{e7.11-2}
p_{b_1b_2} \simeq |c_{b_1b_2}|^2 .
\end{equation} 
Eq.~(\ref{e7.11-2}) is the standard quantum-mechanical formula, containing information 
about all the correlations between the measurements on
two systems, including those that violate Bell inequalities \cite{bell}.

In the derivation above we have not used any explicit analysis of particle trajectories,
so it may not be totally obvious how exactly the nonlocal particle trajectories explain
the final result (\ref{e7.11-2}). 
This is because the main concern of the theory of measurement is the statistics
of positions of the macroscopic pointer, and not the behavior of individual
particle trajectories. Nevertheless, nonlocality and relativity are hidden in
the intermediate step (\ref{e7.5-2}). This equation is relativistic covariant 
and could not be right if we did not have spacetime equivariance (\ref{st_equivariance})
for all times, including times before the probability density approached
the decohered form in the second line of (\ref{e7.5-2}).
And needles to say, the spacetime-equivariance equation 
(\ref{st_equivariance}) itself could not be satisfied if the velocities $V_a^{\mu}$ were not nonlocal.
  
There is another interesting place where the role of relativity can be seen more explicitly
on the observable level. What if the two measuring apparatuses move with respect to each other
with a velocity comparable to the velocity of light? The point is that (\ref{e7.6-2})
is a relativistic scalar {\em in each $y_a$ separately}. This means that we can 
use one system of reference for $y_1$, another system of reference for $y_2$, and still 
obtain the correct result. Thus, instead of (\ref{e7.8}) we will have   
\begin{equation}\label{e7.8-2}
 \rho_{\rm cond}({\bf y}_1,{\bf y}_2';y^0_1,y'^0_2) \simeq 
\sum_{b_1,b_2} |c_{b_1b_2}|^2 \, |\tilde{\phi}_{b_1}({\bf y}_1,y^0_1)|^2|\tilde{\varphi}_{b_2}({\bf y}_2',y'^0_2)|^2 .
\end{equation}
Here the primed and unprimed coordinates denote coordinates calculated in two different systems
of reference, each corresponding to the system in which the corresponding pointer moves
with a velocity much slower than light. It is easy to see that this does not change the 
final result (\ref{e7.11-2}), which implies that
{\em the relative motion of the detectors does not affect nonlocal quantum correlations}.

 \subsection{Operational rules for practical physicists}
\label{SEC7.4}

Now let us reconsider all these results from the point of view of a practical theoretical physicist
who is not interested about any details concerning microscopic particle trajectories
or macroscopic particle detectors. What he wants are the simplest possible 
practical operational rules which will allow him to calculate the probabilities of measurement
outcomes. From our results above such rules can be extracted easily.

So here are the rules. Let the wave function be expanded as
\begin{equation}\label{e7.3r}
\psi(x)=\sum_b c_b \psi_b(x) , 
\end{equation}
where 
\begin{equation}
 \sum_b |c_b|^2=1 ,
\end{equation}
and $\psi_b(x)$ are appropriately normalized wave functions. Let the measurement be such that the
final effective wave function after the measurement will be one of the functions $\psi_b(x)$.
Then the probability that the final effective wave function will be $\psi_b(x)$ is equal to
\begin{equation}\label{e7.11r}
p_b= |c_b|^2 .
\end{equation}
Up to some details, that's all what the practical theoretical physicist needs to 
keep from Sec.~\ref{SEC7.2}. 

One of the details is to specify what does it mean that
$\psi_b(x)$ are ``appropriately normalized''. There are two options. One option is to
use the covariant normalization
\begin{equation}\label{e7normr}
\int d^4x \, |\psi_b(x)|^2 =1.
\end{equation}
Another option is to use the non-covariant normalization 
\begin{equation}\label{e7normr2}
\int d^3x \, |\psi_b({\bf x},t)|^2 =1.
\end{equation}
However, since the wave function is a positive energy wave function, the integral in
(\ref{e7normr2}) does not depend on $t$ (Appendix \ref{SECB.1}), so the two
normalizations differ by a single universal factor $T=\int dt$, which does not affect
the measurable probability (\ref{e7.11r}). Therefore, 
from the point of view of the practical physicist, it doesn't matter which of the 
two normalizations, (\ref{e7normr}) or (\ref{e7normr2}), one will use.

Of course, someone equipped only with the rules above 
(together with their obvious $n$-particle generalization)
may rediscover violation of 
Bell inequalities and other puzzling nonlocal features of QM, and ask: How that can be?
The rules themselves do not provide an answer. The rest of the analysis in
Sec.~\ref{SEC7} does.

\section{Conclusion}
\label{SEC8}
 
The basic motivation for this paper was to completely clarify various conceptual and technical
aspects of relativistic Bohmian mechanics, in the approach which (i) does not need preferred 
foliation of spacetime and (ii) generalizes the usual space probability and space equivariance to
spacetime probability and spacetime equivariance.
With this motivation in mind, as a byproduct we have also clarified various aspects
of the concept of time and probability in more familiar contexts, such as
non-relativistic Bohmian mechanics, classical relativistic mechanics, 
and even classical non-relativistic mechanics. 

On the classical side, we have offered a deeper explanation of the fact that probability
is naturally viewed as a quantity conserved {\em in time}, and shown that in the relativistic
case the time with respect the probability
is naturally conserved is the {\em proper time} along particle trajectories. 
We have used this to explain
why space probability and space equivariance natural in non-relativistic physics
should generalize to spacetime probability and spacetime equivariance
in relativistic physics. We have also explained how, under certain circumstances
not necessarily corresponding to the non-relativistic limit,
space equivariance can be derived from the more fundamental spacetime equivariance.
In addition, we gave a detailed presentation of the many-time formalism 
in classical mechanics and its relation to the more familiar single-time formalism.
Finally, we have presented (Appendix \ref{SECC})
a simple argument that, even in the classical context, 
microscopic proper time may significantly differ from the macroscopic one. 

In the context of non-relativistic Bohmian mechanics 
we have offered a novel qualitative answer to the question why
microscopic Bohmian trajectories cannot be directly observed, and why, at the same time,
the trajectories of macroscopic objects can. For that purpose we have pointed out that
microscopic Bohmian trajectories in QM are analogous to dark matter in astrophysics.

Concerning the main subject of this paper, relativistic Bohmian mechanics, 
first we have identified two fundamental postulates on which this theory is based.
The first postulate is a nonlocal law which determines the fundamental microscopic proper time 
along particle trajectories, while the second is a nonlocal law 
for the particle trajectories themselves. From these postulates 
we have derived the fundamental spacetime-equivariance equation, 
explained how an approximate space-equivariance equation emerges
from the fundamental spacetime-equivariance equation, and gave an argument 
that massless particles do not have Bohmian trajectories.
We have also developed
the relativistic theory of quantum measurement and explained (i) how the spacetime 
probability for the macroscopic pointer leads to the more familiar space probability 
for the macroscopic pointer, and (ii) how that provides the consistency of relativistic
Bohmian mechanics with probabilistic predictions of standard quantum theory. 
In addition (Appendix \ref{SECA}) we have explained how some arguments in the literature 
against existence of relativistic-covariant versions of Bohmian mechanics are circumvented 
by our theory, and argued that nonlocality of the quantum proper time 
cannot be avoided in a wide class of theories obeying
a spacetime-equivariance equation.

\section*{Acknowledgments}
\addcontentsline{toc}{section}{Acknowledgments}

This work was supported by the Ministry of Science of the
Republic of Croatia under Contract No.~098-0982930-2864.

\appendix

\section{Arguments against (naive versions of) relativistic Bohmian mechanics}
  \label{SECA}

 \subsection{A critical overview of existing arguments}
\label{SECA.1}

In the literature one can find various general arguments that a relativistic-covariant version of 
Bohmian mechanics is impossible. Yet in this paper we claim that we have constructed exactly this.
So to fully defend our theory, we must explicitly explain how exactly such arguments
are circumvented by our theory.  

The most frequent arguments of this form have been reviewed in \cite{nikIJQI11}, 
together with explicit explanations of how the arguments are circumvented
by relativistic-covariant Bohmian mechanics. Here we do the same for some 
not-so-frequent arguments, which, however, are proposed by distinguished experts in the field.
We present them in the historical order.

\subsubsection{Argument by Hardy}

In \cite{hardy}, Hardy considers a specific wave function of two entangled particles
and then determines the corresponding Bohmian trajectories by using two different versions of 
non-relativistic Bohmian mechanics, each formulated in a different frame of reference. 
Not surprisingly, he obtains two different results
which contradict each other, from which he concludes that one of the frames must be preferred,
which violates Lorentz invariance.

What is wrong in this argument is an incorrect understanding of the concept of
``violation of Lorentz invariance''. To understand that, one does not need quantum mechanics. 
For example, every macroscopic material object described by classical physics is at rest
with respect to some particular reference frame, so by a similar kind of reasoning one might
conclude that such an object also ``violates Lorentz invariance''.
But this of course is wrong. Lorentz invariance is not a property of one particular object,
or of one particular set of Bohmian trajectories. Lorentz invariance is a property 
of the {\em general law} that determines the motion of this object or particles.
But the general law itself does not determine the motion uniquely, because motion
depends also on some initial conditions. The initial condition, of course, cannot be 
Lorentz invariant, which is why a specific solution of the Lorentz invariant equations
may define some preferred frame. A preferred frame obtained in this way, however,
does not violate Lorentz invariance of the theory.          

So in the case Hardy considers, the relativistic covariant law for particle trajectories
(which does {\em not} have the form of non-relativistic Bohmian mechanics in one particular
Lorentz frame) has many solutions. One of the solutions may resemble the trajectories of the first
version of non-relativistic Bohmian mechanics, while another solution may resemble 
the trajectories of the second version of non-relativistic Bohmian mechanics.
Which one will be realized in a given run of the Hardy experiment depends on initial
conditions of hidden variables (i.e. spacetime particle positions at an initial proper time $s$)
in that run. In this way, for each run one may have a different preferred frame without 
violation of Lorentz invariance.
 
\subsubsection{Argument by Berndl {\it et al}}

In \cite{berndl96}, Berndl, D\"urr, Goldstein and Zangh\`i consider a relativistic covariant law
for Bohmian particle trajectories similar to that in the present paper. For such trajectories they show 
that $|\psi|^2$ cannot be the {\em space} probability density in all Lorentz frames and that
these trajectories do not lead to {\em space} equivariance (which they simply call -- equivariance).
They use this result to argue that with such a theory it is difficult to make concrete
statistical predictions, which makes difficult to show that the theory is compatible
with the predictions of standard QM.

What we have shown in this paper is that (i) the statistical predictions still {\em can} be 
made, and (ii) that they {\em do agree} with standard QM. We were able to do that because we have used a property
which they did not consider -- the spacetime equivariance instead of space equivariance.
The spacetime equivariance gives the spacetime probability density (not the space probability
density), but we have explained how under certain conditions, especially those which correspond
to measurement by a macroscopic apparatus, space probability density can be obtained 
from the spacetime probability density.  

\subsubsection{Argument by Valentini}

In \cite{valentini}, Valentini makes the point that 
the fundamental laws of non-relativistic Bohmian mechanics
determine the {\em velocity} of the particle, unlike those of classical
mechanics which in the Newton form determine the {\em acceleration} of the particle.
From that he concludes that non-relativistic Bohmian mechanics is not
Galilean invariant. Then, by viewing Lorentz transformations as a modification
of Galilean transformations, he argues that it is not reasonable to expect that 
relativistic Bohmian mechanics should be Lorentz invariant.

Our response is that viewing Lorentz transformations as a modification
of Galilean transformations is not the only way how the Lorentz transformations
can be viewed. Another (in fact, more modern) view, originally introduced by
Minkowski, is to view Lorentz transformations as a generalization of space 
rotations. In the language of group theory, space rotations are elements
of the group SO(3), while Lorentz transformations are spacetime rotations, i.e. elements
of the group SO(1,3). 
This, indeed, is how we view Lorentz transformations in this paper.
Non-relativistic Bohmian mechanics is space-rotation invariant,
and in the same sense in which non-relativistic Bohmian mechanics is space-rotation invariant, 
relativistic Bohmian mechanics is Lorentz invariant.

\subsubsection{Argument by Gisin} 

\begin{figure}[t]
\centerline{\includegraphics[width=8cm]{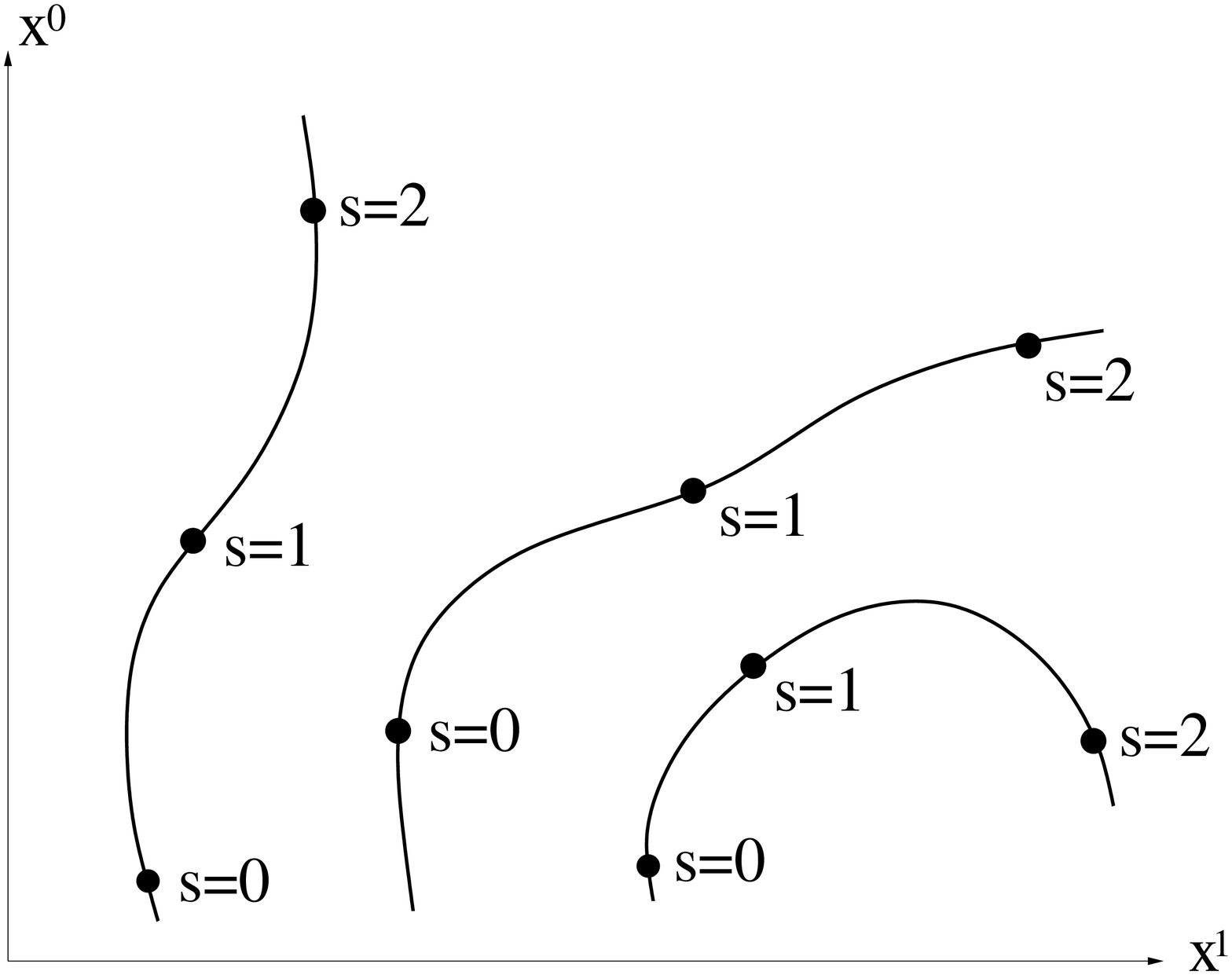}}
\caption{\label{fig0} An example of relativistic Bohmian trajectories for a system with 3 particles.}
\end{figure}

In \cite{gisin}, Gisin shows that any nonlocal hidden variable theory compatible with the predictions of
QM cannot be relativistic covariant, in the sense that the time ordering of events cannot be invariant
under Lorentz transformations.

Our main objection to this argument is the abuse of terminology. Namely, in relativistic literature,
the property of time-ordering invariance is not called ``covariance''. Instead, Lorentz covariance
is a property that the general law of motion has the same form in all Lorentz frames, 
which relativistic Bohmian mechanics obeys.

Otherwise, this general conclusion by Gisin is fully satisfied in relativistic Bohmian mechanics.
We illustrate this in Fig.~\ref{fig0}, where we show an example of Bohmian trajectories for a system
of 3 particles. The fundamental microscopic time ordering of hidden variables (particle positions in spacetime)
is defined by the value of quantum proper time $s$;
those with larger value of $s$ come ``later''.
It is clear from the picture that this $s$-ordering does not coincide with the ordering
defined by an arbitrary Lorentz frame. In fact, in this example the $s$-ordering does not coincide 
with the ordering defined by {\em any} Lorentz frame. 

Fig.~\ref{fig0} also nicely illustrates why in relativistic Bohmian mechanics
there is no preferred foliation of spacetime. In an attempt to define a preferred hypersurface
which connects all points with the same value of $s$, one immediately recognizes that there is an
infinite number of such hypersurfaces for any given value of $s$. Neither of them
is ``preferred'' in any meaningful sense. 

In addition, those who are worried that some particles 
in Fig.~\ref{fig0} move faster than light (which necessarily implies that in some Lorentz frames
they move backwards in time) should see Sec.~\ref{SEC7.3.3}.

 \subsection{A new conjecture: Quantum proper time cannot be local}
\label{SECA.2}

In (\ref{e5.14}) we have postulated a nonlocal law that determines the  quantum proper time. 
Later we have shown that such a nonlocal
proper time is compatible with predictions of QM. We do not claim that (\ref{e5.14}) is the only possible
law compatible with predictions of QM. Therefore we address the following question: Is it possible to replace 
(\ref{e5.14}) by a {\em local} law for the quantum proper time? In this subsection we argue that it is
not possible. Or more precisely, we propose the following conjecture:  
\newtheorem{conjecture}{Conjecture}
\begin{conjecture}
Let particle trajectories satisfy an equation of the form
\begin{equation}\label{A1}
 \frac{dX_a^{\mu}}{d\lambda}=U_a^{\mu}, 
\end{equation}
where $\lambda$ is the quantum proper time. Then, in general, the spacetime-equivariance equation
\begin{equation}\label{A2}
\frac{\partial|\psi|^2}{\partial \lambda} + \sum_{a=1}^{n} \partial_{a\mu}(|\psi|^2 U_a^{\mu}) =0 
\end{equation}
can only be satisfied if the law for $\lambda$ is nonlocal.
\end{conjecture}

We do not have a rigorous proof of the conjecture. Nevertheless, we can present a simple argument for its validity
as follows. Eq.~(\ref{A1}) implies 
\begin{equation}\label{A7}
 d\lambda^2 =\frac{1}{U_{a\nu}U_{a}^{\nu}} dX_{a\mu}dX_a^{\mu} .
\end{equation}
The requirement that this law for $\lambda$ should be local means that
\begin{equation}\label{A8}
 U_{a\mu}U_{a}^{\mu}=l_a(x_a) ,
\end{equation}
where $l_a(x_a)$ is some function which does not depend on other positions except $x_a$.
This suggests that $U_{a}^{\mu}$ should also be a local function $U_{a}^{\mu}(x_a)$, 
but this seems impossible to reconcile with the equivariance equation (\ref{A2})
for an arbitrary nonlocal wave function $\psi(x_1,\ldots,x_n)$. 
Therefore, it seems that (\ref{A8}) cannot be true, implying that the law
(\ref{A7}) cannot be local.

\section{Positive-energy wave functions}
\label{SECB}

In this appendix we study some properties of positive-energy wave functions.
For simplicity we study the case of 1-particle wave function, but all the results 
can be trivially generalized to $n$-particle wave functions as well.

 \subsection{Inner product and normalization}
\label{SECB.1}

Consider a wave packet
\begin{equation}\label{eB2.1}
 \psi(x) = \int \frac{d^3p}{(2\pi)^3} \, c({\bf p}) e^{-ip\cdot x} ,
\end{equation}
where the energy
\begin{equation}
p^0=E({\bf p})=\sqrt{{\bf p}^2+m^2}
\end{equation}
is positive. If energy is positive in one Lorentz frame, then $m^2\geq 0$ implies that it is positive
in all Lorentz frames.
The measure $d^3p$ is not Lorentz invariant, but the transformation of $c({\bf p})$
under Lorentz transformations 
can be defined such that (\ref{eB2.1}) is Lorentz invariant.
This can be seen by writing
\begin{equation}
 d^3p\, c({\bf p}) =\frac{d^3p}{E({\bf p})}\, E({\bf p}) c({\bf p}) 
\equiv \frac{d^3p}{E({\bf p})}\, \tilde{c}({\bf p}) ,
\end{equation}
because the measure $d^3p/E({\bf p})$ is Lorentz invariant, so (\ref{eB2.1}) is Lorentz invariant
if we define the transformation of $c({\bf p})$ such that $\tilde{c}({\bf p})$ transforms
as a Lorentz scalar. 

Now consider the inner product
\begin{equation}\label{inner}
 \langle \psi'| \psi\rangle = \int d^3x\, \psi'^*(x) \psi(x) ,
\end{equation}
where $\psi'(x)$ is another positive-energy wave packet with $c'({\bf p})$ instead of $c({\bf p})$.
Since $d^3x$ is not a Lorentz-invariant measure, this inner product is not Lorentz invariant.
Nevertheless, this inner product is interesting even in the relativistic context 
because it does not depend on $x^0$. Here we prove this by generalizing a result in \cite{nikFP08}.
Using the expansion (\ref{eB2.1}) for both $\psi$ and $\psi'$ and exploiting the identity
\begin{equation}
 \int \frac{d^3x}{(2\pi)^3} \, e^{i({\bf p}-{\bf p}'){\bf x}} =\delta^3({\bf p}-{\bf p}'), 
\end{equation}
(\ref{inner}) gives
\begin{equation}\label{inner2}
 \langle \psi'|\psi\rangle = \int \frac{d^3p'}{(2\pi)^3} \int d^3p \,
 c'^*({\bf p}') c({\bf p}) e^{-i(E({\bf p})-E({\bf p}')) x^0} \delta^3({\bf p}-{\bf p}') .
\end{equation}
The factor $\delta^3({\bf p}-{\bf p}')$ implies that the only contribution to the integral 
comes from ${\bf p}={\bf p}'$, in which case $E({\bf p})-E({\bf p}')=0$. This implies
that $e^{-i(E({\bf p})-E({\bf p}')) x^0}=1$, so (\ref{inner2}) reduces to
\begin{equation}\label{inner3}
 \langle \psi'|\psi\rangle = \int \frac{d^3p}{(2\pi)^3} \,
 c'^*({\bf p}) c({\bf p}) ,
\end{equation}
which does not depend on $x^0$. In this derivation, the crucial assumption is that both wave
packets are positive-energy wave packets. If we had a superposition of both positive and 
negative energies, then we would have factors of the form
$e^{-i(E({\bf p})+E({\bf p}')) x^0}$ or $e^{i(E({\bf p})+E({\bf p}')) x^0}$,
which would not be $x^0$-independent when 
${\bf p}={\bf p}'$. Of course, if we had only negative energies, the inner product would be
$x^0$-independent again. 

As a special case, (\ref{inner3}) implies that the norm
\begin{equation}\label{norm}
  \langle \psi |\psi\rangle = \int d^3x\, \psi^*(x) \psi(x) = 
\int \frac{d^3p}{(2\pi)^3} \, |c({\bf p})|^2 
\end{equation}
does not depend on $x^0$. In this paper we use this result several times.

 \subsection{Average values and related properties}  
\label{SECB.2}

For a given wave function $\psi$ and a given operator $\hat O$, we define the quantity
\begin{equation}
 \langle O \rangle = \frac{\langle \psi |\hat O |\psi\rangle }{\langle \psi |\psi\rangle} ,
\end{equation}
and refer to it as ``average value'', having in mind
that this may not have the usual statistical meaning of average value if
$\psi^*\psi$ cannot be interpreted as a probability density in space.
Likewise, we define the ``standard deviation''
\begin{equation}
\Delta O = \sqrt{ \langle O^2 \rangle - \langle O \rangle^2 } , 
\end{equation}
with a similar caveat regarding its statistical meaning. 

A quantity of a particular interest is the average energy, which can be expressed as 
\begin{equation}
 \langle E \rangle = \langle \psi |\psi\rangle^{-1} \int d^3x \, \psi^*(x) i\partial_0 \psi(x) . 
\end{equation}
It is easy to see that this is equal to 
\begin{equation}
 \langle E \rangle =  \langle \psi |\psi\rangle^{-1} 
\int\frac{d^3p}{(2\pi)^3} \, |c({\bf p})|^2 E({\bf p}) , 
\end{equation}
which does not depend on $x^0$.

In the following we shall consider more complicated
quantities involving products of $\psi^*$ and $\psi$. To save writing,
for that purpose it is convenient to introduce the short-hand notation
\begin{equation}
 e^{-i(p-p')\cdot x} \equiv z(p,p',x) ,
\end{equation}
\begin{equation}
 \int \frac{d^3p'}{(2\pi)^3} \int \frac{d^3p}{(2\pi)^3} \, c^*({\bf p}') c({\bf p}) \, F(p,p',x)
 \equiv \int F,
\end{equation}
where $F(p,p',x)$ is an arbitrary function.
With this notation we find 
\begin{equation}
 |\psi|^2=\int z ,
\end{equation}
\begin{equation}
 j_0= \frac{i}{2m} \, \psi^* \stackrel{\leftrightarrow\;}{\partial_{0}} \psi 
= \int \frac{p_0+p_0'}{2m} \, z ,
\end{equation}
so
\begin{equation}
 V^0=\frac{j^0}{\psi^*\psi}=\frac{\displaystyle\int \frac{p_0+p_0'}{2m} \, z}{\displaystyle\int z} .
\end{equation}
From those results we find
\begin{equation}
 V^0 \partial_0 |\psi|^2 = \frac{ \displaystyle\int \frac{p_0+p_0'}{2m} \, z \displaystyle\int \partial_0 z }
{\displaystyle\int z} , 
\end{equation}
\begin{equation}
 |\psi|^2  \partial_0 V^0 = \frac{ \displaystyle\int z \displaystyle\int \frac{p_0+p_0'}{2m} \, \partial_0 z
- \displaystyle\int \frac{p_0+p_0'}{2m} \, z \displaystyle\int \partial_0 z }
{\displaystyle\int z} ,
\end{equation}
so we can write
\begin{equation}\label{eB2.e1}
 |\psi|^2 \partial_0 V^0 = V^0 \partial_0 |\psi|^2 \, \epsilon ,
\end{equation}
where
\begin{equation}\label{eB2.e2}
 \epsilon=\frac{ \displaystyle\int z  \displaystyle\int \frac{p_0+p_0'}{2m} \, \partial_0 z }
{\displaystyle\int \partial_0 z \displaystyle\int \frac{p_0+p_0'}{2m} \, z } -1.
\end{equation}

Now let us assume that the wave packet (\ref{eB2.1}) is very narrow in the $k_0$ space, i.e. that
the uncertainty of energy $\Delta E$ is much smaller than the average energy  $\langle E \rangle \equiv \bar{E}$. 
In that case the integrals in (\ref{eB2.e2}) can be approximated by replacing 
the integration variables $p_0$ and $p_0'$ with their average value $\bar{E}$. The average value
is a constant that can be put in front of the integral, so (\ref{eB2.e2}) becomes
\begin{equation}\label{eB2.e3}
 \epsilon\simeq \frac{ \displaystyle\frac{\bar{E}}{m}\displaystyle\int z  \displaystyle\int \, \partial_0 z }
{\displaystyle\frac{\bar{E}}{m} \displaystyle\int \partial_0 z \displaystyle\int  \, z } -1=0.
\end{equation}
This shows that actually
\begin{equation}\label{eB2.e4}
 \epsilon\ll 1 ,
\end{equation}
so (\ref{eB2.e1}) implies 
\begin{equation}\label{eB2.e5}
 |\psi|^2 \partial_0 V^0 \ll V^0 \partial_0 |\psi|^2 .
\end{equation}
This justifies the approximation
\begin{equation}\label{eB2.e6}
 \partial_0 (|\psi|^2 V^0)=V^0 \partial_0 |\psi|^2 + |\psi|^2 \partial_0 V^0 \simeq V^0 \partial_0 |\psi|^2
\end{equation}
used in (\ref{e5.40}).
   
\section{Macroscopic vs microscopic proper time in classical mechanics}
\label{SECC}

\begin{figure}[t]
\centerline{\includegraphics[width=6cm]{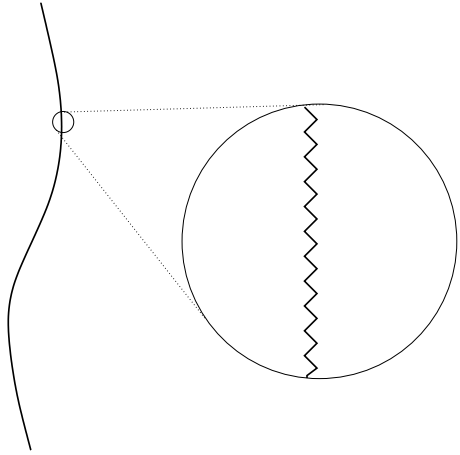}}
\caption{\label{fig1} {\it Left}: Spacetime trajectory on the macroscopic level,
with velocity fairly smaller than the velocity of light.
{\it Right}: Zoomed spacetime trajectory on the microscopic level,
having the velocity of light everywhere
except on a set of measure zero.}
\end{figure}

To understand that macroscopic proper time may differ from the microscopic one,
one does not necessarily need relativistic Bohmian mechanics. 
Such a phenomenon appears also at the purely classical level.

For that purpose consider the spacetime trajectory of a macroscopic object in 
Fig.~\ref{fig1}. It is a trajectory of an object moving relatively slowly
compared to the velocity of light, so the proper time $\tau$
associated with the trajectory is quite large.

On the other hand, by zooming the trajectory down to microscopic 
distances, in Fig.~\ref{fig1} one reveals that the trajectory is actually a zigzag
trajectory, such that the actual velocity is equal to the
velocity of light everywhere except on a set of measure zero.
So the microscopic proper time $\tau$ is zero.

So, what is the true proper time associated with the trajectory?
This question is very much analogous to the question
``How long is the coast of Britain?''
famously asked by Mandelbrot 
to explain the idea of fractals (see e.g. \cite{chaos}). 
The answer is that there is no ``true''
length. Instead,
the length depends on scale on which the measurement is performed.

Analogously, there is no ``true''
proper time either. 
It depends on the scale on which the measurement is performed. 
In the case of Fig.~\ref{fig1}, a microscopic clock will 
measure zero proper time, while a macroscopic clock will
measure a fairly large proper time. There should be 
nothing mysterious about that.

\end{document}